\documentclass[aps,prb,twocolumn,superscriptaddress,floatfix,citeautoscript,hyperref=pdftex]{revtex4-1}
\usepackage{times}
\usepackage{graphicx}
\usepackage{dcolumn}
\usepackage{bm}
\usepackage{color}
\usepackage{amsmath}
\usepackage{amssymb}
\newcommand {\beq} {\begin{equation}}
\newcommand {\eeq} {\end{equation}}
\newcommand {\bqa} {\begin{eqnarray}}
\newcommand {\eqa} {\end{eqnarray}}

\usepackage[colorlinks=true]{hyperref}

\begin{document}

\title{The phase diagram of the underdoped cuprates at high magnetic field}

\author{Debmalya Chakraborty}
\affiliation{Institut de Physique Th\'eorique, CEA, Universit\'e Paris-Saclay, Saclay, France}

\author{Corentin Morice}
\affiliation{Institut de Physique Th\'eorique, CEA, Universit\'e Paris-Saclay, Saclay, France}

\author{Catherine P\'epin}
\affiliation{Institut de Physique Th\'eorique, CEA, Universit\'e Paris-Saclay, Saclay, France}

\begin{abstract}

The experimentally measured phase diagram of cuprate superconductors in the temperature-applied magnetic field plane illuminates key issues in understanding the physics of these materials. At low temperature, the superconducting state gives way to a long-range charge order with increasing magnetic field; both the orders coexist in a small intermediate region. The charge order transition is strikingly insensitive to temperature, and quickly reaches a transition temperature close to the zero-field  superconducting $T_c$. We argue that such a transition along with the presence of the coexisting phase cannot be described simply by a competing orders formalism. We demonstrate that for some range of parameters there is an enlarged symmetry of the strongly coupled charge and superconducting orders in the system depending on their relative masses and the coupling strength of the two orders. We establish that this sharp switch from the superconducting phase to the charge order phase can be understood in the framework of a composite SU(2) order parameter comprising the charge and superconducting orders. Finally, we illustrate that there is a possibility of the coexisting phase of the competing charge and superconducting orders only when the SU(2) symmetry between them is weakly broken due to biquadratic terms in the free energy. The relation of this sharp transition to the proximity to the pseudogap quantum critical doping is also discussed. 

\end{abstract}

\maketitle

\section{Introduction}\label{sec:Intro}

Enigmatic high transition temperature superconductors, especially cuprates, are ideal playgrounds to understand various facets of condensed matter physics. The presence of the superconducting phase prevents the investigation of the underlying normal state properties at low temperatures. The application of an external magnetic field suppresses superconductivity giving a novel way out to explore the normal state. We argue that the phase diagram of cuprates in the temperature (T)-applied magnetic field (B) plane can enlighten us with some key aspects of these superconductors.  

Underdoped cuprates display a mysterious pseudogap phase\cite{Alloul89,Alloul91,Warren89}, where the antinodal regions of the Brillouin zone are gapped out\cite{Campuzano98, Vishik:2012cc, Vishik12, Yoshida:2012kh, He14, Vishik14}. This phenomenon occurs at higher temperatures than the superconducting transition temperature ($T_c$). While the formation of the pseudogap phase can be associated with $\vec{Q}=0$ (translational symmetry preserving) orders like loop currents\cite{Fauque06, ManginThro:2014js, ManginThro:2015fg} or nematicity\cite{Mesaros:2011jj,Sato2017}, the exact origin and nature of this pseudogap phase remains to be completely understood. Over the years, numerous experiments \cite{Hucker11,Tranquada95,PhysRevB.89.224513,Comin14,Tabis14,daSilvaNeto14,Ghiringhelli12,Chang12,Achkar12,Wu:2015bt,Wu13a,Wu11,Thampy13,Hamidian15a} revealed the ubiquitous existence of a $\vec{Q} \ne 0$ charge density wave order in the pseudogap phase of the underdoped cuprates. The study of this charge order can significantly help in the understanding of the puzzling pseudogap phase. 

X-ray scattering measurements \cite{Ghiringhelli12,Blackburn13a,Blackburn13b,Blanco-Canosa13,Chang2010} in ${\rm Y}{\rm Ba}_2{\rm Cu}_3{\rm O}_{y}$ (YBCO) at zero magnetic field identified the existence of charge density wave modulations in the doping range $0.09 \le p \le 0.13$. The correlation lengths of this charge order were found to be $\sim 20$ lattice spacings along the ${\rm Cu}{\rm O}_2$ planes and $\sim 1$ lattice spacing in the perpendicular direction. This established the two dimensional (2D) and short-range nature of these modulations. The in-plane modulations showed incommensurate bidirectional checkerboard patterns with the dominant wave vectors being $\vec{Q} \approx (0.3,0)$ and $\vec{Q} \approx (0,0.3)$. Similar checkerboard charge modulations were also observed in Bi-based cuprates using scanning tunneling microscopy \cite{Wise08,HowaldKapitulnik03} at low temperatures. The onset temperature ($T_{co}^0$) of this short-range charge order in YBCO was found to be much higher than $T_c$.

At high magnetic fields, nuclear magnetic resonance (NMR) line splittings \cite{Wu:2015bt,Wu13a,Wu11} showed the presence of charge modulations in YBCO for $p \sim 0.11-0.12$. This was further supported by sound velocity measurements \cite{LeBoeuf13}, which indicated a thermodynamic phase transition to a true long-range charge order (CO) at an onset field $B_{co}$ ($\sim$ 17 T) for $p=0.11$. In the same doping range ($\sim 0.11-0.12$), quantum oscillations \cite{Doiron-Leyraud07, Sebastian12, Barisic2013} associated with a negative Hall \cite{LeBoeuf07, LeBoeuf11, Doiron-Leyraud13, 2015arXiv150805486G} constant and a negative Seebeck \cite{Laliberte11, Doiron-Leyraud13, Chang2010} coefficient point towards the formation of a small electron pocket in the Fermi surface at high magnetic fields ($B > 25$ T). The reconstruction of the Fermi surface from large hole arcs at high doping to small electron pockets at low doping is attributed to the broken translational symmetry due to the presence of the charge modulations of a substantial range. Whether the modulations responsible for this Fermi surface reconstruction correspond to the bidirectional checkerboard patterns or unidirectional stripe patterns is still under debate. Charge modulations obtained in high-field X-ray scattering measurements \cite{Chang16,Gerber:2015gx,Jang14645} also show profound signatures. The in-plane correlation lengths of these modulations become as high as $\sim 100$ lattice spacings, confirming the long-range nature of the high-field CO. Additionally, these high-field X-ray measurements suggest the presence of an incommensurate unidirectional three dimensional (3D) CO with out-of-plane correlation length $\sim 10$ lattice spacings. The 3D CO has the same in-plane incommensuration as the 2D counterpart. This indicates that the appearance of the 3D CO is somehow related to the 2D long-range CO. At high magnetic fields, all these experiments thus demonstrate the appearance of a long-range CO irrespective of its structure. 

A competition between the CO and the superconductivity (SC) is already noticeable at zero or moderate magnetic fields. First evidence of this competition can be viewed from the suppression of the zero field $T_c$ in the same doping range where the short-range CO is observed. Second, the intensity of the zero-field X-ray scattering CO peaks decreases for $T<T_c$. The presence of a moderate magnetic field reduces this decrease. The competition can be further substantiated by scanning tunneling microscopy, which detects the short-range CO in regions of space where the amplitude of the superconductivity is reduced (both near the vortex core \cite{Hoffman02} at low fields and surrounding ${\rm Zn}$ impurities \cite{2016Natur.532..343H} at zero magnetic field). 

\begin{figure}[t]
\includegraphics[width=7.4cm,height=6.0cm]{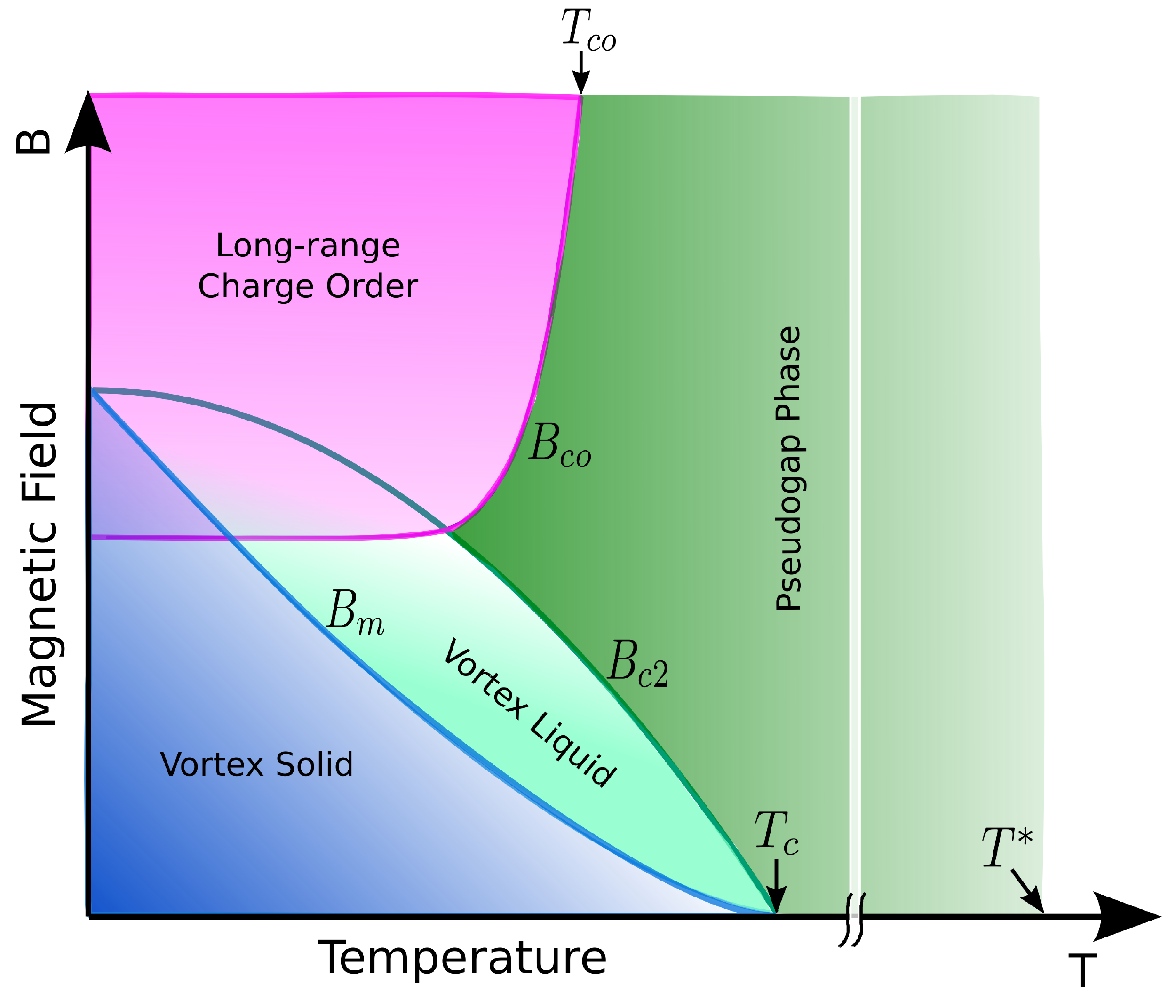}
\caption{ A schematic B-T phase diagram of underdoped YBCO summarizing various experiments. Type-II superconductors have two critical fields: a lower critical field ($B_{c1}$) and a upper critical field ($B_{c2}$). The system completely expels magnetic fields for $B<B_{c1}$ showing the Meissner effect and allows magnetic field flux lines to penetrate at various locations (called vortices) for $B_{c1}<B<B_{c2}$. Cuprates have very low $B_{c1}$ and so form vortices with the application of a very small magnetic field. $B_{c2}$ varies with temperature as shown in the figure and its exact profile depends on the specifics of the sample. At low temperatures, the vortices form periodic arrays called vortex solid with local short-range charge modulations inside the vortex core. Increasing the temperature, this solid melts for $B>B_m$. Though $B_m$ has a different temperature dependence than $B_{c2}$, it intersects the $B_{c2}$ line in the two different limits of zero temperature and zero magnetic field. At high magnetic field, system shows a long-range charge density wave order with the transition field $B_{co}$. $B_{co}$ is insensitive to temperature at low temperatures and marks a sudden rise. $T_{co}$ is the temperature at high magnetic fields where the charge order marks a transition to the pseudogap phase. $T_{co}$ is effectively insensitive to the magnetic field. The green region is the pseudogap phase where one looses any coherence of either charge or superconducting order. The pseudogap phase persists for temperatures below $T^*$, which is very large compared to $T_c$ or $T_{co}$. At low temperatures, the magenta charge order region merges with the blue superconducting order region showing the coexistence of both the orders. The short-range charge order is present even for low magnetic fields, but is not shown in this schematic.        
} 
\label{fig:expschematic}
\end{figure}

In this paper, we will focus on the competition between the SC and the long-range CO at high magnetic fields. This competition is prominent in the B-T phase diagram of underdoped cuprates. In Fig.~\ref{fig:expschematic}, we show a schematic B-T phase diagram summarizing various experiments on underdoped ($0.11 \le p \le 0.13$) YBCO. Our endeavor in this work will be to understand the following salient features of the long-range CO in the B-T phase diagram:      

\begin{enumerate}

\item The long-range CO phase is associated with an onset magnetic field ($B_{co}$) which is found to be surprisingly insensitive to the temperature and remains flat up to the scale of $T_{c}$ \cite{2014NatCo...5E3280G,LeBoeuf13}. 

\item At high magnetic fields, the long-range CO transition temperature ($T_{co}$) is nearly independent of the magnetic field \cite{LeBoeuf13}. Remarkably, sound velocity measurements \cite{LeBoeuf13}, NMR \cite{Wu13a} and thermal conductivity measurements \cite{2014NatCo...5E3280G} all suggest that $T_{co}$ is very close to $T_c$ and significantly lower than $T_{co}^0$ (transition temperature of the short-range CO).

\item The sound velocity measurements \cite{LeBoeuf13} indicated that there exists a coexisting phase in the B-T phase diagram at low temperatures. This was also manifested in the thermal conductivity measurements \cite{2014NatCo...5E3280G} at low temperatures, which identified the onset field of the long-range CO to be lower than the upper critical field of the superconductor.  

\end{enumerate}     
In particular, we argue that the flatness of $B_{co}$ is a signature of an SU(2) symmetry between the SC and the CO. The coexisting phase in the phase diagram is a result of a weak biquadratic symmetry breaking between the two, caused by interaction terms in the free energy.

Theoretically, competing orders\cite{RevModPhys.87.457,Kivelson:2002er,Demler:2001hp,PhysRevB.80.035117,Zhang:2002hz,Benfatto00,Caprara:2016gs,PhysRevB.95.224511,PhysRevLett.75.4650,PhysRevB.54.16216,Wang15a,Wang15b} are studied enormously in the context of the underdoped cuprates. In the presence of a magnetic field, the SC is suppressed near a vortex core. As a result, any competing order like charge density wave \cite{Einenkel14,Kivelson:2002er,Zhang:2002hz}, spin density wave \cite{Demler:2001hp,PhysRevB.80.035117,Zhang:2002hz,PhysRevB.66.214502} or pair density wave \cite{PhysRevB.91.104512} becomes recognizable near the vortex cores. The competing orders are often treated within a Ginzburg-Landau theory. In particular, it was shown in Ref.~\onlinecite{Kivelson:2002er} that the CO can coexist with the SC in a halo surrounding the vortex core where the SC is suppressed partially. This CO inside each halo fluctuates enormously with no long-range CO. It was postulated that only a interlayer coupling or a finite magnetic field (inducing vortex-vortex interaction) can stabilize a true long-range CO.

In the first part of this paper (Sec.~\ref{sec:GL}), we focus on a similar Ginzburg-Landau theory of the competing SC and 2D CO, but in a different perspective. We will treat a Ginzburg-Landau free energy with effective homogeneous order parameters (averaging the vortex induced inhomogeneities) near the upper critical magnetic field of the superconductor. In this approach, the magnetic field renormalizes the effective mass (coefficient of the quadratic term of order parameters in the free energy) of the SC order parameter. We couple this effective free energy of the SC with the free energy of the CO and study the competition. Note that we only consider the long-range CO. We phenomenologically construct the B-T phase diagram with increasing coupling strength between the two orders. This helps us quantifying the relation between the region of the coexisting phase and different parameters in the Ginzburg-Landau theory (like coupling strength and the mass of the two order parameters). We show that a strong competition between the SC and the CO leads to a phase diagram with no coexisting phase. Within this mean field picture, we infer that the temperature insensitivity of $B_{co}$ is specific to an extreme fine tuning of the temperature dependence of each mass parameter. We demonstrate that, for a range of parameters, there is an enlarged symmetry between the SC and the CO where they are energetically degenerate. For this regime of parameters, the associated massless fluctuations \cite{Zhang97,Efetov13} of the two order parameters become important and cannot be captured in a Ginzburg-Landau picture. 

A similar enhanced symmetry between the SC and the CO is proposed in Ref. \onlinecite{Efetov13} in the pseudogap phase of the underdoped cuprates. In this approach, the pseudogap phase is characterized by a composite SU(2) order parameter comprising the SC and the CO. The SU(2) symmetry between these suborders imposes a constraint on them, reflecting their strong competition. Fluctuations associated with this symmetry are described by a non linear sigma model \cite{Efetov13}. This SU(2) theory is successful in describing some of the phenomenological aspects \cite{Meier13,Montiel:2016it,PhysRevB.93.024515,PhysRevB.96.094529,PhysRevB.96.134511,PhysRevB.95.104510} of the much debated pseudogap phase. A similar non linear sigma model describing the fluctuating CO and SC was also studied in Ref.~\onlinecite{Hayward14,PhysRevB.92.224504} which explain many trends of the zero or low field X-ray scattering data.

In the second part of this paper (Sec.~\ref{sec:NLSMintro}), we study the competition of the SC and the CO within the SU(2) theory. We use a renormalization group treatment of the associated nonlinear sigma model (similar to the one developed in Ref. \onlinecite{Meier13}) and illustrate the B-T phase diagram. We show that the temperature insensitivity of $B_{co}$ at low temperatures is a unique feature of the SU(2) theory. Our analysis shows that $T_{co} \approx T_c$, another exclusive feature of the SU(2) theory. We discuss the role of the underlying SU(2) symmetry in the pseudogap phase, which is characterized as a disordered phase of the fluctuating SC or CO. We further illustrate the possibility of the presence of a coexisting phase even in the existence of a strong constraint between the SC and the CO. 

We also predict some features of the B-T phase diagram in the doping range $0.13 \le p \le 0.2$ (close to the pseudogap quantum critical doping under the superconducting dome). Using a quantum non linear sigma model \cite{Chakravarty89}, we postulate that the width of the flatness of $B_{co}$ reduces as the doping goes close to the pseudogap quantum critical doping. Near this critical doping, $T_{co}$ at high magnetic field is no longer independent of the magnetic field at which the measurement is done.

\section{Ginzburg-Landau theory of competing orders}\label{sec:GL}

\begin{figure*}[t]
\includegraphics[width=0.9\textwidth]{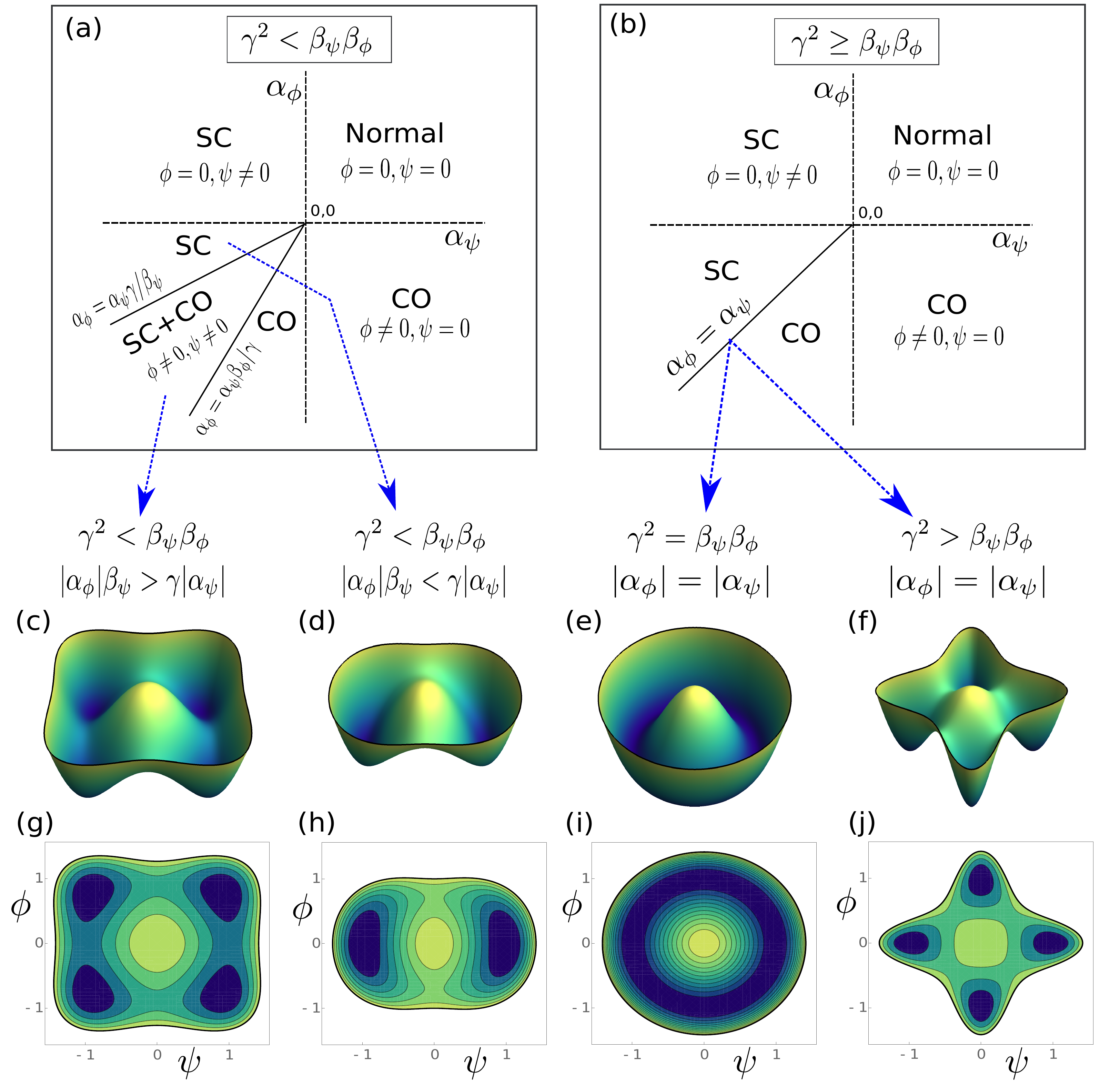}
\caption{ An illustration of the four different phases possible in a GL free energy (Eq.~\eqref{eq:GLf1}) with two competing orders (here, superconducting and charge order). If $\alpha_{\psi}$ (mass of the superconducting field) and $\alpha_{\phi}$ (mass of the charge order field) are both positive, only the normal state is stable; if $\alpha_{\psi}<0$ and $\alpha_{\phi}>0$, only the superconducting (SC) phase is stable; if $\alpha_{\psi}>0$ and $\alpha_{\phi}<0$, only the charge order (CO) phase is stable; if $\alpha_{\psi}<0$ and $\alpha_{\phi}<0$, there is a possibility of a coexisting (SC+CO) phase. If the repulsion strength ($\gamma$) between the fields is small ($\gamma^2<\beta_{\psi} \beta_{\phi}$), the phase diagram in (a) accommodates the SC+CO phase in the region bound by two lines given by the mass conditions in Eqs.~\eqref{eq:Con2} and \eqref{eq:Con3}. Strong $\gamma$ disfavors any coexistence of the fields as shown in (b). We show the free energy density landscapes (c-f) and their contour maps (g-j) in the plane of the order parameters for four different GL parameter regimes. (c,g): If the conditions for the coexistence are satisfied, we see four minima in the free energy density corresponding to both $\psi \neq 0$ and $\phi \neq 0$. (d,h): Even if $\gamma$ is small, the free energy density will form minima in either $\psi=0$ or $\phi=0$ sectors depending on the mass conditions in Eq.~\eqref{eq:Con2} and \eqref{eq:Con3}. Here, we show the case where Eq.~\eqref{eq:Con3} is violated. (e,i): For the specific case of $\gamma^2=\beta_{\psi} \beta_{\phi}$ and $\alpha_{\psi}=\alpha_{\phi}$, the free energy density shows an enlarged symmetry with no change in its value for a fixed $\psi^2+\phi^2$. The Mexican hat shape of the free energy density is evident. (f,j): $\gamma^2 > \beta_{\psi} \beta_{\phi}$ disfavors any coexistence with minima in both $\psi=0$ and $\phi=0$ axes. In this case, the stable phase is governed only by $|\alpha_{\psi}|$-$|\alpha_{\phi}|$ (case with $|\alpha_{\psi}|=|\alpha_{\phi}|$ is shown in the plot (f) and (j)). In all the plots, we have taken $\beta_{\psi}=\beta_{\phi}=1$.   
} 
\label{fig:freeenergy}
\end{figure*}

\subsection{Generic features of the free energy: conditions for coexistence}\label{sec:GL1}

Ginzburg-Landau (GL) theories are used extensively to describe phase transitions phenomenologically without getting into the microscopic details of a system. The main idea behind this formulation is to write the free energy density in powers of the order parameters corresponding to broken symmetries near the transition. The GL free energy describing systems with multiple broken symmetries can be written as a sum of the free energies for each broken symmetry if there is no interaction between the fields describing individual order parameters. The competition or repulsion between the fields increases this free energy. This imposes a restriction on the strength of the interaction for the existence of a coexisting phase. In the following, we derive the conditions imposed on this interaction strength based on the GL theory of two competing order parameters: the superconducting and the charge order parameters. The free energy density functional of two order parameters, $\psi$ (describing the SC order) and $\phi$ (describing the CO) is given by:
\begin{equation}
f[\psi,\phi]=\alpha_{\psi} \psi^2 + \frac{\beta_{\psi}}{2} \psi^4 + \alpha_{\phi} \phi^2 + \frac{\beta_{\phi}}{2} \phi^4 + \gamma \psi^2 \phi^2
\label{eq:GLf1}
\end{equation}
where $\psi$ and $\phi$ are $N_1$ and $N_2$ component fields respectively, $\gamma$ is the coupling between the two fields and $\beta_{\psi},\beta_{\phi}>0$. In Eq.~\eqref{eq:GLf1}, we have kept terms up to the fourth order in fields. Our calculation in this section is for general $N_1$ and $N_2$, unless mentioned. 

In the absence of any coupling between the two fields, both the fields condense to form a state with $\psi \neq 0$ and $\phi \neq 0$ if $\alpha_{\psi}<0$ and $\alpha_{\phi}<0$. In the presence of the coupling between the fields, there exists four possible phases: the SC phase ($\psi \neq 0$ and $\phi = 0$), the CO phase ($\phi \neq 0$ and $\psi = 0$), the coexisting phase ($\psi \neq 0$ and $\phi \neq 0$) and a normal state ($\phi = 0$ and $\psi = 0$). An illustration of these phases is shown in Fig. \ref{fig:freeenergy}(a) and (b). The mean field solution of Eq.~\eqref{eq:GLf1} can be obtained by minimizing the free energy with respect to the order parameters $\psi$ and $\phi$. For the SC phase and the CO phase, we have the solutions $\psi^2=-\alpha_{\psi}/\beta_{\psi}$ and $\phi^2=-\alpha_{\phi}/\beta_{\phi}$ respectively. When both the orders coexist, we can obtain the solution by minimizing the free energy simultaneously with respect to $\psi$ and $\phi$, which yields:
\begin{equation}
\alpha_{\psi} + \beta_{\psi} \psi^2 + \gamma \phi^2=0
\label{eq:equations1}
\end{equation}
\begin{equation}
{\rm and}~~~~ \alpha_{\phi} + \beta_{\phi} \phi^2 + \gamma \psi^2=0
\label{eq:equations2}
\end{equation}
These coupled equations have a unique solution for $ \gamma^2 \neq \beta_{\psi}\beta_{\phi}$:
\begin{equation}
\psi^2=\frac{\gamma \alpha_{\phi}-\alpha_{\psi} \beta_{\phi}}{\beta_{\psi}\beta_{\phi}-\gamma^2},\phi^2=\frac{\gamma \alpha_{\psi}-\alpha_{\phi} \beta_{\psi}}{\beta_{\psi}\beta_{\phi}-\gamma^2}
\label{eq:Solpsiphi}
\end{equation}
The mean field free energy density for the SC phase and the CO phase is $f_{\rm sc}=-\alpha_{\psi}^2/(2 \beta_{\psi})$ and $f_{\rm co}=-\alpha_{\phi}^2/(2 \beta_{\phi})$ respectively. The free energy corresponding to the coexisting phase is given by:
\begin{equation}
f_{\rm sc+co}=f_{\rm sc}-\frac{\left ( \alpha_{\phi}\beta_{\psi}-\gamma \alpha_{\psi} \right )^2}{2\beta_{\psi} \left ( \beta_{\psi}\beta_{\phi}-\gamma^2 \right )}=f_{\rm co}-\frac{\left ( \alpha_{\psi}\beta_{\phi}-\gamma \alpha_{\phi} \right )^2}{2\beta_{\phi} \left ( \beta_{\psi}\beta_{\phi}-\gamma^2 \right )}
\label{eq:mffreeIII}
\end{equation} 
From Eq.~\eqref{eq:mffreeIII}, it is clear that the coexisting phase can be stable with respect to either the SC phase or the CO phase if $f_{\rm sc+co}<f_{\rm sc}$ and $f_{\rm sc+co}<f_{\rm co}$, which gives: 
\begin{equation}
\gamma^2 < \beta_{\psi}\beta_{\phi}
\label{eq:Con1}
\end{equation}
The coexisting phase also demands the existence of a solution $\psi \neq 0$ and $\phi \neq 0$ in Eq.~\eqref{eq:Solpsiphi}. This gives two more conditions on the masses:
\begin{equation}
\gamma \alpha_{\phi}>\alpha_{\psi} \beta_{\phi}
\label{eq:Con2}
\end{equation}
\begin{equation}
{\rm and}~~~~ \gamma \alpha_{\psi}>\alpha_{\phi} \beta_{\psi}
\label{eq:Con3}
\end{equation}
The conditions in Eqs.~\eqref{eq:Con1}, \eqref{eq:Con2} and \eqref{eq:Con3} can only be satisfied with $\alpha_{\psi}<0$ and $\alpha_{\phi}<0$, which is a necessary but not sufficient condition. With both masses ($\alpha_{\psi}$ and $\alpha_{\phi}$) being negative, conditions in Eqs.~\eqref{eq:Con2} and \eqref{eq:Con3} can also be re-written as $\gamma \left |\alpha_{\phi} \right |<\left |\alpha_{\psi} \right | \beta_{\phi}$ and $\gamma \left |\alpha_{\psi} \right |<\left |\alpha_{\phi} \right | \beta_{\psi}$. 

We plot the free energy density ($f$) in Fig.~\ref{fig:freeenergy} for $\beta_{\psi}=\beta_{\phi}=1$. We first discuss the case when the coupling between the fields is weak enough satisfying condition in Eq.~\eqref{eq:Con1}. Additionally, if we satisfy both the conditions in Eqs.~\eqref{eq:Con2} and \eqref{eq:Con3}, the coexisting phase becomes stable as shown in Fig.~\ref{fig:freeenergy}(c) and (g). If one of these conditions is not satisfied, the coexisting phase is not stable any more and the positions of the minima of $f$ shift to either $\phi=0$ if $\left | \alpha_{\phi} \right | < \gamma \left | \alpha_{\psi} \right |$ (shown in Fig.~\ref{fig:freeenergy}(d) and (h)) or $\psi=0$ if $\left | \alpha_{\phi} \right |>\left | \alpha_{\psi} \right | /\gamma$.  

If we increase the coupling between the fields such that we satisfy $\gamma^2 = \beta_{\psi}\beta_{\phi}$, the system of Eqs.~\eqref{eq:equations1} and \eqref{eq:equations2} has no unique solution. Hence, we show that this coupling is special and results into an enhanced symmetry in the free energy density. Indeed, if we scale the masses and the fields by the corresponding coefficients of their quartic potentials such that $\bar{\alpha}_{\psi}=\alpha_{\psi}/\sqrt{\beta_{\psi}}$, $\bar{\alpha}_{\phi}=\alpha_{\phi}/\sqrt{\beta_{\phi}}$, $\bar{\psi}^2=\psi^2/\sqrt{\beta_{\psi}}$ and $\bar{\phi}^2=\phi^2/\sqrt{\beta_{\phi}}$, we can rewrite the system of Eqs.~\eqref{eq:equations1} and \eqref{eq:equations2} as:
\begin{equation}
\bar{\alpha}_{\psi} + \bar{\psi}^2 + \bar{\phi}^2=0
\label{eq:equationsscaled1}
\end{equation}
\begin{equation}
{\rm and}~~~ \bar{\alpha}_{\phi} + \bar{\phi}^2 + \bar{\psi}^2=0
\label{eq:equationsscaled2}
\end{equation}
If these scaled masses of the two fields are the same ($\bar{\alpha}_{\psi}=\bar{\alpha}_{\phi}$), we have a larger symmetry between $\psi$ and $\phi$ fields: the free energy is invariant if we keep $\bar{\psi}^2+\bar{\phi}^2$ fixed. The two fields are degenerate with no energy cost needed to rotate from one to the other. In this case the free energy, which is $O(N_1)\times O(N_2)$ symmetric in general, displays a higher symmetry of $O(N_1+N_2)$. This $O(N_1+N_2)$ symmetry is visible in the Mexican hat like form of the free energy in Fig.~\ref{fig:freeenergy}(e) and (i). The constraint of fixed $\bar{\psi}^2+\bar{\phi}^2$ introduces fluctuations in each $\bar{\psi}$ and $\bar{\phi}$. These fluctuations can be treated within a $O(N_1+N_2)$ non linear sigma model (see Sec.~\ref{sec:NLSMintro} for details). 

Further increasing $\gamma^2$ above $\beta_{\psi} \beta_{\phi}$, pushes the minima in the free energy to either the SC phase or the CO phase depending on their relative masses. If $\alpha_{\psi}=\alpha_{\phi}$, all the four minima are degenerate as shown in Fig.~\ref{fig:freeenergy}(f) and (j).    

\subsection{Free energy in the presence of an external magnetic field}\label{sec:GL2}

In a type-II superconductor, the external magnetic field does not penetrate the sample below a lower critical field $B_{c1}$ due to the Meissner effect. If the magnetic field (B) is increased above $B_{c1}$, the magnetic field couples to the orbital motion of the electrons and the flux lines penetrate the sample through different locations creating vortices. This state is commonly known as the mixed phase. The magnitude of the SC order parameter vanishes at the core of these vortices. The inhomogeneities arising due to the vortices will add gradient terms in the free energy functional of the superconductor which is given by:
\begin{equation}
F_{\rm sc}-F_{\rm n}=\int \alpha'_{\psi} \psi^{2}(r) + \frac{\beta_{\psi}}{2} \psi^4(r) + \frac{\lambda}{2} \left | \left ( \frac{\nabla}{i} - \frac{2e\vec{A}}{c} \right ) \psi(r) \right |^2 dR
\label{eq:GLgrad1}
\end{equation} 
where $F_{\rm sc}$ is the free energy functional of the superconductor alone, $F_{\rm n}$ is the free energy functional of the normal state and $\vec{A}$ is the vector potential corresponding to the magnetic field. Cuprates are commonly known as extremely type-II superconductors with a high Ginzburg-Landau parameter (which is the ratio of the penetration depth and the coherence length of the superconductor). As a result, these superconductors have a very small $B_{c1}$ and there is effectively no screening of magnetic field by Meissner currents, i.e., $\nabla \times \vec{A}=B \hat{z}$, where $B$ is the external applied magnetic field. We choose $z$ as the direction perpendicular to the orbital motion of the electrons in the 2D ${\rm Cu}{\rm O}_2$ planes of the superconductor. As the magnetic field is further increased, the number of vortices increases and their separation decreases. There exists an upper critical magnetic field $B_{c2}$ where the order parameter collapses resulting in a second order transition to the normal phase. Close to $B_{c2}$, the SC order parameter $\psi$ is small and the free energy density can be treated (see appendix \ref{sec:appendixHC2}) within an effective homogeneous theory. In terms of an average order parameter $\psi$, the free energy density of the superconductor is written as:  
\begin{equation}
f_{\rm sc}-f_{\rm n}=\alpha_{\psi} \psi^{2} + \frac{{\beta_{\psi}}}{2} {\psi}^4
\label{eq:GLgrad2}
\end{equation}
where the mass term $\alpha_{\psi}$ is renormalized due to magnetic field and is given by:
\begin{equation}
\alpha_{\psi}=\alpha'_{\psi}+\zeta B+a_{sc}T^2 
\label{eq:alpha1}
\end{equation}
with $\alpha'_{\psi}<0$ and $\zeta$ is a positive constant. We take a quadratic temperature dependence of $\alpha_{\psi}$ as we are expanding near zero temperature. Near the transition temperature, the temperature dependence of $\alpha_{\psi}$ can be well approximated as linear in $T$. $a_{sc}$ is the measure of the tolerance of the superconducting order to thermal suppression. The mass term $\alpha_{\psi}$ changes its sign when the magnetic field reaches its upper critical value:
\begin{equation}
B_{c2}=(\alpha'_{\psi}+a_{sc}T^2)/\zeta
\label{eq:uppercrit}
\end{equation}

\begin{figure}[t]
\includegraphics[width=0.45\textwidth]{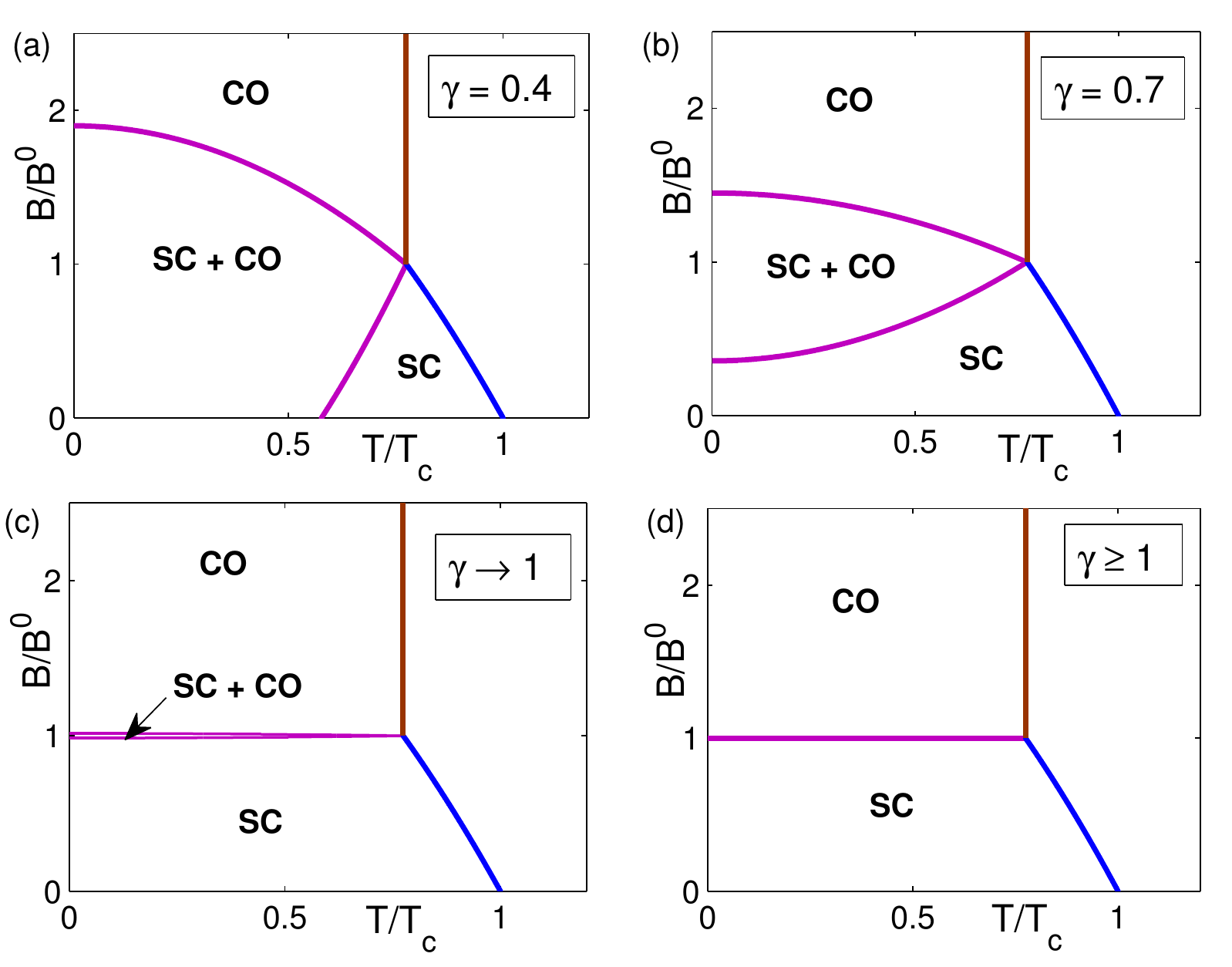}
\caption{ The applied magnetic field-temperature phase diagram in a GL theory with competing $\psi$ and $\phi$ orders for different strengths of competition. At low $\gamma$, a coexisting phase with both CO and SC orders is stable at low magnetic fields and temperatures. With increased $\gamma$, the fields repel each other strongly to reduce the coexisting region with its disappearance at $\gamma=1$. At $\gamma=1$, the two second order magenta lines (given by Eqs.~\eqref{eq:Bound1} and~\eqref{eq:Bound2}) merge to form a first order line, as shown in (d). The first order magenta line has an enlarged symmetry of $O(N_1+N_2)$ with $\alpha_{\psi}=\alpha_{\phi}$. The topology of the phase diagram remains same for all values of $\gamma \geq 1$. We have scaled the magnetic field with $B^0$ and temperature with $T_{c}$ and used $a_{sc}=a_{co}=1$. The first order magenta line $B=B^0$ in (d) is independent of temperature primarily because we choose $a_{sc}=a_{co}=1$ (see Eq.~\eqref{eq:Bound3new}). We also choose $\beta_{\psi}=\beta_{\phi}=1$.
}
\label{fig:BT1}
\end{figure}

We can now include the form of the SC free energy in Eq.~\eqref{eq:GLgrad2} in our free energy functional for the coupled SC and CO system:
\begin{equation}
f[\psi,\phi]=f_{\rm sc}-f_{\rm n} + \alpha_{\phi} \phi^2 + \frac{\beta_{\phi}}{2} \phi^4 + \gamma \psi^2 \phi^2
\label{eq:GLgrad3}
\end{equation}
where $\alpha_{\phi}$ is parametrized as:
\begin{equation}
\alpha_{\phi}=\alpha'_{\phi}+a_{co}T^2 
\label{eq:alpha2}
\end{equation}
with $\alpha'_{\phi}<0$. $a_{co}$ is the measure of the thermal suppression of the CO order parameter. We neglect the temperature dependence of $\beta_{\psi}$ and $\beta_{\phi}$. The form of Eq.~\eqref{eq:GLgrad3} is the same as in Eq.~\eqref{eq:GLf1}, but in Eq.~\eqref{eq:GLgrad3}, $\psi$ or $\phi$ are the effective homogeneous order parameters and $\alpha_{\psi}$ is the renormalized SC mass. 

\subsubsection*{\textbf{\textit{B-T phase diagram}}}

We now use the free energy density in Eq.~\eqref{eq:GLgrad3} to construct the B-T phase diagram. As illustrated earlier in Sec.~\ref{sec:GL1}, we can access a coexisting SC and CO phase if the repulsion between the SC and CO fields is weak enough.  
In this regime of weak interaction between the fields ($\gamma^2<\beta_{\psi} \beta_{\phi}$), the coexisting phase is restricted to a region in the B-T phase diagram bounded by two lines obtained from the conditions in Eqs.~\eqref{eq:Con2} and \eqref{eq:Con3}. The boundary line separating the SC phase and the coexisting phase is given by:
\begin{equation}
B_{{\rm sc} \rightarrow {\rm sc+co}}(T)=\frac{1}{\zeta} \left[ \left( \frac{\gamma\alpha'_{\phi}}{\beta_{\phi}} -\alpha'_{\psi} \right) + \left( \frac{\gamma a_{co}}{\beta_{\phi}}-a_{sc} \right) T^2 \right]
\label{eq:Bound1}
\end{equation} 
and the boundary line separating the coexisting phase and the CO phase is given by:
\begin{equation}
B_{{\rm sc+co} \rightarrow {\rm co}}(T)=\frac{1}{\zeta} \left[ \left( \frac{\beta_{\psi} \alpha'_{\phi}}{\gamma} -\alpha'_{\psi} \right) + \left( \frac{\beta_{\psi} a_{co}}{\gamma}-a_{sc} \right) T^2 \right]
\label{eq:Bound2}
\end{equation}
If $\gamma^2=\beta_{\psi} \beta_{\phi}$, the two lines in Eqs.~\ref{eq:Bound1} and~\ref{eq:Bound2} merge to form a single line. If $\beta_{\psi}=\beta_{\phi}=\gamma$, this single line reduces to:
\begin{equation}
B_{{\rm sc} \rightarrow {\rm co}}(T)=\frac{1}{\zeta} \left[ \left( \alpha'_{\phi} -\alpha'_{\psi} \right) + \left( a_{co}-a_{sc} \right) T^2 \right]
\label{eq:Bound3new}
\end{equation}  
For the analysis of the phase diagram, we choose $\alpha'_{\psi}=-1$, $\alpha'_{\phi}=-0.6$, $\beta_{\psi}=\beta_{\phi}=1$, $\zeta=1$. 

In Fig.~\ref{fig:BT1}, we plot the B-T phase diagram corresponding to the free energy in Eq.~\eqref{eq:GLgrad3} with increasing coupling strength between the fields for $a_{sc}=a_{co}=1$. The magnetic field lines (blue lines in Fig.~\ref{fig:BT1}) marking the transition from the SC phase to the normal phase is given by the condition $\alpha_{\psi}=0$ in Eq.~\eqref{eq:alpha1}, which yields $B(T)=(1/\zeta)(-\alpha'_{\psi}-a_{sc}T^2)$. At $B=0$, $\alpha_{\psi}=0$ gives the transition temperature ($T_{c}$) as:
\begin{equation}
T_{c}=\sqrt{\frac{-\alpha'_{\psi}}{a_{sc}}}
\label{eq:tc1}
\end{equation}
The transition from the CO phase to the normal state at high magnetic field is independent of the magnetic field. This transition (brown line in Fig.~\ref{fig:BT1}) is given by the condition $\alpha_{\phi}=0$ in Eq.~\eqref{eq:alpha2} and occurs at a temperature $T_{co}$ given by:
\begin{equation}
T_{co}=\sqrt{\frac{-\alpha'_{\phi}}{a_{co}}} 
\label{eq:tc2}
\end{equation}
For $\gamma<1$ (regime of weak repulsion), the coexisting phase is stable in a region of the phase diagram bounded by two lines (magenta lines in Fig.~\ref{fig:BT1}) given by expressions in Eqs.~\eqref{eq:Bound1} and \eqref{eq:Bound2}. The magenta lines meet the blue lines (characterizing the transition from the SC phase to the normal phase) and the brown lines (characterizing the transition from the CO phase to the normal phase) at a multicritical point ($T_{co}$, $B^0$), where $B^0=|\alpha'_{\psi}|-|\alpha'_{\phi}|$. If $\gamma<|\alpha'_{\phi}|/|\alpha'_{\psi}|$, the coexisting phase is stable even at $B=0$ (Fig. \ref{fig:BT1}(a)). Increasing $\gamma$ shrinks the region of coexistence with eventual overlap of the two magenta lines at $\gamma=1$. 

As explained in Sec.~\ref{sec:GL1}, the free energy has an enlarged symmetry for $\gamma^2=\beta_{\psi}\beta_{\phi}$. For our choice of parameters ($\beta_{\psi}=\beta_{\phi}=1$) in this section, the condition for the enhanced symmetry reduces to $\gamma=1$. If $\gamma=1$, there is no coexistence of the SC and the CO phase. The transition magnetic field line from the SC phase to the CO phase is governed by Eq.~\eqref{eq:Bound3new} and is shown by the magenta line in Fig.~\ref{fig:BT1}(d). The individual masses $\alpha_{\psi}$ and $\alpha_{\phi}$ become equal along this magenta line. For $B<B^0$, the mass of the SC field is smaller than the mass of the CO field ($\alpha_{\psi}<\alpha_{\phi}$). This stabilizes only the SC phase. The mass of the SC field increases with increasing magnetic field and becomes equal to the mass of the CO field at $B=B^0$. For $B>B^0$, the mass of the SC field becomes greater than the mass of the CO field. Consequently, the CO phase gets stabilized for $B>B^0$. Therefore, when $\gamma=1$, the transition from the SC phase to the CO phase is decided by the individual masses of each of the fields. If we further strengthen $\gamma$, the stable phase is still governed by the size of the individual masses only. The topology of the B-T phase diagram remains same for all $\gamma \geq 1$. The effect of the competition between the fields is visible only for $T<T_{co}$. The transitions from the SC to the normal phase and from the CO phase to the normal phase are independent of the coupling strength. So, the brown lines and the blue lines in Fig.~\ref{fig:BT1} are at the same place for all $\gamma$. $T_{co}$ and $T_{c}$ are the temperatures where the individual masses ($\alpha_{\psi}$ and $\alpha_{\phi}$) vanish and are not connected to each other in general. 

\begin{figure}[t]
\includegraphics[width=0.45\textwidth]{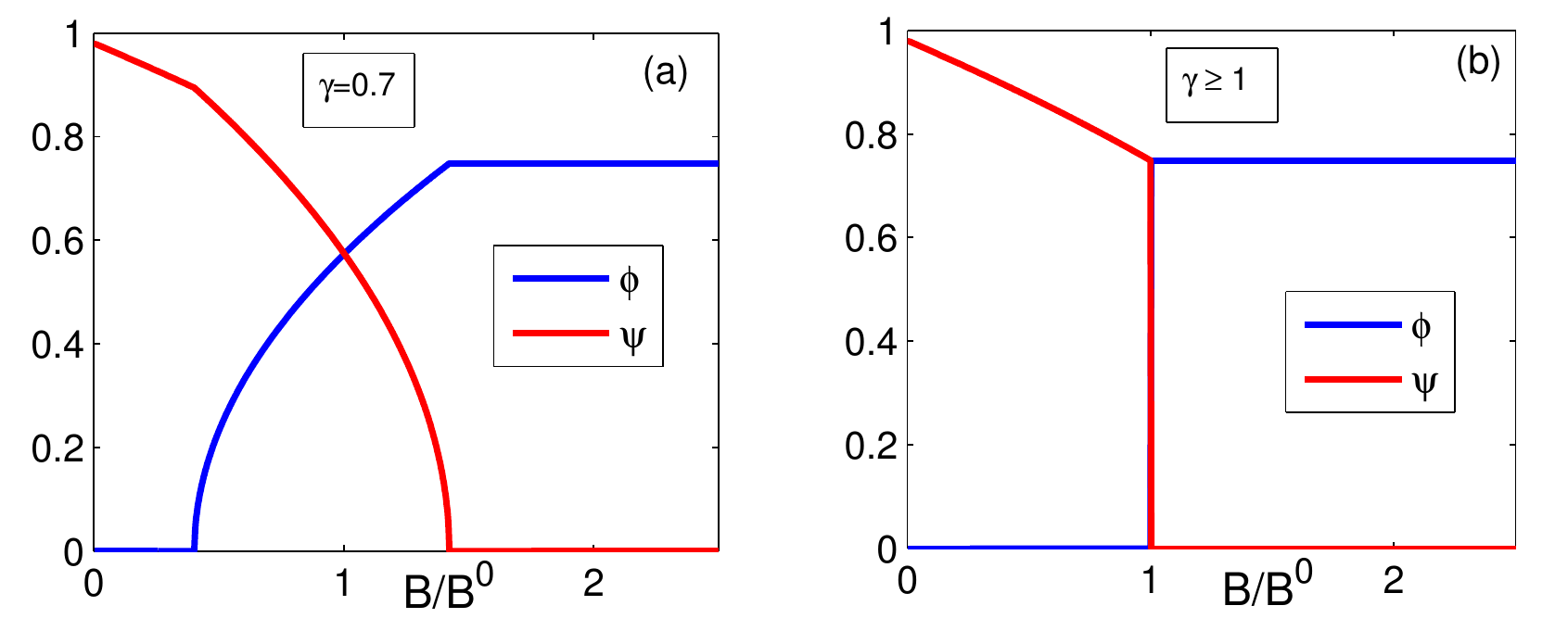}
\caption{ The order parameters $\psi$ and $\phi$ are plotted using the expressions in Eq.~\eqref{eq:Solpsiphi} for the GL free energy in Eq.~\eqref{eq:GLgrad3}. They are shown as a function of $B/B^0$ at $T=0.2T_{c}$ for $\gamma=0.7$ (a) and $\gamma \geq 1$ (b). Both $\psi$ and $\phi$ behave as a continuous function for $\gamma<1$ indicating that the transitions from the SC phase to the coexisting phase to the CO phase are all second order in nature. The coexistence region ceases to exist for $\gamma \geq 1$. At $B=B^0$, $\psi$, $\phi$ and consequently $f$ experience discontinuous changes marking a first order transition as shown in (b).      
}
\label{fig:opB}
\end{figure}

The transitions from the SC phase or the CO phase to the normal state are second order transitions for all $\gamma$ as the order parameters $\psi$ or $\phi$ vanish continuously at the transition lines. In contrast, the nature of the transition lines $B_{{\rm sc} \rightarrow {\rm sc+co}}$ or $B_{{\rm sc+co} \rightarrow {\rm co}}$ for $\gamma<1$ in Fig.~\ref{fig:BT1}(a-c) and the transition line $B_{{\rm sc} \rightarrow {\rm co}}$ in Fig.~\ref{fig:BT1}(d) are completely different. Transition lines from the SC phase to the coexisting phase and from the coexisting phase to the CO phase correspond to second order transitions. But the transition from the SC phase to the CO phase is a first order transition for $\gamma \geq 1$ as the order parameters experience a discontinuous jump at $B=B^0$. We show the profile of the order parameters $\psi$ and $\phi$ as a function of magnetic field in Fig.~\ref{fig:opB} for $\gamma<1$ (a) and $\gamma \ge 1$ (b).
The first order transition from the SC to the CO has not been reported in experiments \cite{Chang16}. 

\subsubsection*{\textbf{\textit{Different temperature dependence of the individual masses}}}

\begin{figure}[t]
\includegraphics[width=0.45\textwidth]{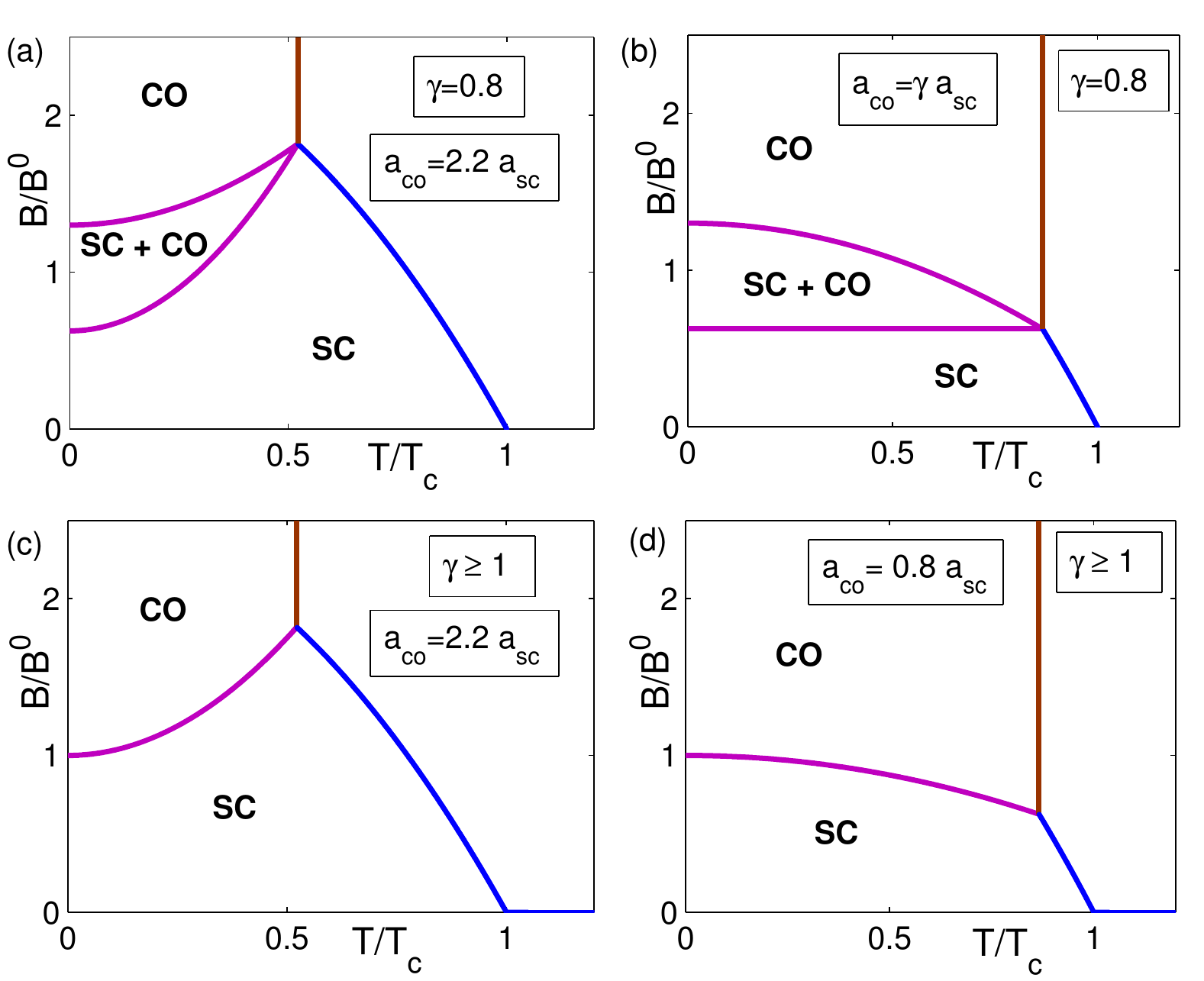}
\caption{ The B-T phase diagrams for the GL free energy in Eq.~\eqref{eq:GLgrad3} with $a_{sc} \neq a_{co}$. The flat magenta line in Fig.~\ref{fig:BT1}(d) is reminiscent of the fact that we take $a_{sc}=a_{co}$ in Eq.~\eqref{eq:Bound3new}. The slope of first order transition lines (magenta) in (c,d) depend crucially on the values of $a_{sc}$ (parameter deciding the temperature dependence of $\alpha_{\psi}$) and $a_{co}$ (parameter deciding the temperature dependence of $\alpha_{\psi}$). The first order lines are flat if the temperature dependence of the masses of both the fields are exactly same. The second order transition lines ($\gamma<1$) are shown in (a) and (b). The transition from the SC phase to the coexisting phase is flat only if $a_{co}=\gamma a_{sc}$. This figure reiterates the fact that the flat transition to the CO phase seen in experiments are not generic to a competing order GL theory.
}
\label{fig:BT2}
\end{figure}

The first order transition line demarcating the SC phase and the CO phase is flat with temperature independent $B^0$ in Fig.~\ref{fig:BT1}(d). Looking at this, we might get tempted to interpret it as the temperature independence of the transition field in experiments. This feature is however not true in general and is applicable only for $a_{sc}=a_{co}$. From Eq.~\eqref{eq:Bound3new}, we can see that $B_{{\rm sc} \rightarrow {\rm co}}=B^0$ is independent of $T$ if $a_{sc}=a_{co}$. In Fig.~\ref{fig:BT2}, we explore the phase diagram with $a_{sc} \neq a_{co}$. In the presence of the coexistence, $B_{{\rm sc} \rightarrow {\rm sc+co}}$ in Eq.~\eqref{eq:Bound1} is independent of temperature only if $a_{co}=\gamma a_{sc}/\beta_{\psi}$. For $\gamma \geq 1$, the first order line is only temperature independent if $a_{sc}=a_{co}$. It should be noted that in a phenomenological mean field treatment, $a_{sc}$ and $a_{co}$ are only parameters which determine the temperature dependence of the individual masses. Thus, the flatness of the transition of the SC phase to the CO phase can be achieved in a GL theory by extreme fine tuning of the parameters and is not a generic feature.  

\subsection{Renormalization group approach}\label{sec:GL4}

Cuprates are considered to be short coherence length superconductors and thus have a large Ginzburg region where the mean field theory described above is supposed to fail\cite{chaikin2000principles}. The fluctuations near the critical lines can be captured in a renormalization group (RG) treatment. Within an RG treatment, the bare parameters ($\alpha_{\psi}$, $\alpha_{\phi}$, $\beta_{\psi}$, $\beta_{\phi}$ and $\gamma$) in Eq.~\eqref{eq:GLf1} get renormalized to effective parameters ($\tilde{\alpha}_{\psi}$, $\tilde{\alpha}_{\phi}$, $\tilde{\beta}_{\psi}$, $\tilde{\beta}_{\phi}$ and $\tilde{\gamma}$). The RG analysis of the free energy in Eq.~\eqref{eq:GLf1} is already carried out in Ref.~\onlinecite{PhysRevB.13.412} and we do not replicate the same. However, we summarize the results in the following paragraph.

The RG technique is based on integrating out the fast momentum fluctuations iteratively to write the effective free energy which captures the slow momentum or long wavelength fluctuations. During the process of the iteration, the parameters get modified recursively following trajectories governed by the RG equations. The RG equations have fixed points (FP) which correspond to the scale invariant parameters. These FP can describe a phase or a phase transition depending on their stability. The RG equations corresponding to Eq.~\eqref{eq:GLf1} have six FP:
\begin{enumerate}
\item Trivial FP: $\alpha_{\psi}^*=0$, $\alpha_{\phi}^*=0$, $\gamma^*=0$
\item Gaussian-Heisenberg FP: $\beta_{\psi}^*=0$, $\beta_{\phi}^*\neq 0$, $\gamma^*=0$
\item Heisenberg-Gaussian FP: $\beta_{\psi}^*\neq 0$, $\beta_{\phi}^*=0$, $\gamma^*=0$
\item Heisenberg-Heisenberg FP: $\beta_{\psi}^*\neq 0$, $\beta_{\phi}^*\neq 0$, $\gamma^*=0$
\item First order FP: $\beta_{\psi}^*=\beta_{\phi}^*=\gamma^*$, $\alpha_{\psi}^*=\alpha_{\phi}^*$
\item Second order FP: $\beta_{\psi}^* \neq \beta_{\phi}^* \neq \gamma^*$, $\alpha_{\psi}^* \neq \alpha_{\phi}^*$
\end{enumerate}
The first four FP give $\gamma^*=0$ where the SC order and the CO are decoupled. They represent the transition from the SC phase to the normal state and the transition from the CO phase to the normal state. The fifth FP corresponds to the situation of the enhanced symmetry of $O(N_1+N_2)$ where the effective free energy landscape looks similar to Fig.~\ref{fig:freeenergy}(e) and (i). This FP describes the first order transition from the SC phase to the CO phase. The sixth FP satisfies the mean field criterion for coexistence ($\gamma^{*2}<\beta_{\psi}^*\beta_{\phi}^*$) and represent the free energy landscape similar to Fig.~\ref{fig:freeenergy}(c) and (g). The bare parameters depend on the applied magnetic field ($B$) and temperature ($T$). The transition lines in the B-T phase diagram can be determined by studying the stability of the FP. The stability of the FP depends crucially on the values of $N_1$, $N_2$, $N_1+N_2$ and the dimension of the system. There are several analytical and numerical studies of the stability of these FP in three dimensions \cite{PhysRevB.67.054505,doi:10.1143/JPSJ.69.2395,PhysRevE.88.042141,Demler04}. But the stability of the FP in the case of two dimensions\cite{PhysRevB.74.064407,PhysRevB.76.024436,PhysRevB.82.144531,PhysRevE.94.042105} is more complex and is still an open question. So, it is difficult to pinpoint whether the B-T phase diagram obtained in a competing order formalism can include a coexisting phase or not. 

Moreover, in 2D, the amplitude fluctuations play no role in deciding the critical behavior. Instead, the thermal phase fluctuations, captured by the renormalizations of the gradient terms, are important in deciding the phase boundaries \cite{Jaefari:2010ga,altland2006condensed}. The RG approach discussed in the preceding paragraph treats only the amplitude renormalizations and does not take care of the renormalizations of the gradient terms. Hence, the temperature dependence in the B-T phase diagrams found from the analysis of Eq.~\eqref{eq:GLf1} is not expected to give the correct trends in two spatial dimensions.  

\vskip 0.5cm
In this section, we described the GL theory of the competing superconducting and charge orders in the presence of a magnetic field. We constructed the B-T phase diagram for different strengths of the competition and discussed the possibility of explaining the experimentally observed features. As evident from Fig. \ref{fig:BT1}, strengthening the competition between the SC and the CO fields disfavors any coexisting phase in the phase diagram. $B_{{\rm sc} \rightarrow {\rm sc+co}}$ for $\gamma^2<\beta_{\psi}\beta_{\phi}$ and $B_{{\rm sc} \rightarrow {\rm co}}$ for $\gamma^2 \ge \beta_{\psi}\beta_{\phi}$ are flat only if the temperature dependences of $\alpha_{\psi}$ and $\alpha_{\phi}$ are extremely fine tuned (Fig. \ref{fig:BT2}). Further, the similarity of the $T_c$ at zero field and $T_{co}$ at high field cannot be established in this picture (see Eqs.~\eqref{eq:tc1} and~\eqref{eq:tc2}). These features make us believe that the B-T phase diagram of underdoped cuprates is hard to explain within a GL theory of competing orders.

In Sec.~\ref{sec:GL1}, we could identify a parameter regime where the free energy shows an enlarged $O(N_1+N_2)$ symmetry. The enlarged symmetry puts a constraint (Eq.~\ref{eq:equationsscaled1}) on the SC and the CO fields if the two orders are energetically degenerate. We now turn our discussion to an emergent SU(2) theory where the strongly competing SC order and the CO are nearly degenerate in energy. In the next section, we will first introduce this SU(2) theory and then construct the B-T phase diagram using a renormalization group treatment.

\section{SU(2) symmetry between CO and SC: Non linear sigma model}\label{sec:NLSMintro}

Underdoped cuprates \cite{Metlitski10a,Efetov13} are often described by a two dimensional spin-fermion model \cite{Abanov00,abanov03}. This model features the pseudogap phase \cite{Efetov13} characterizing an emergent SU(2) symmetry connecting a d-wave superconductor and a quadrupole density wave. This quadrupole density wave corresponds to charge density modulations \cite{Meier14,Einenkel14,Montiel2017,Montiel2017sr} in the 2D ${\rm Cu}{\rm O}_2$ plane. The wave vector ($\vec{Q}$) of this CO is typically incommensurate and is taken to be momentum dependent \cite{Montiel2017}. $\vec{Q}$ can therefore correspond to both a unidirectional stripe-like charge order and a bidirectional checkerboard charge order \cite{Montiel2017,Montiel2017sr}. In this section, we focus on the SU(2) symmetry between the SC and the 2D CO, without going into the details of the directionality of the CO. This theory though has broader applicability in describing the symmetry of the CO. We expect that the presence of an interlayer coupling between the 2D ${\rm Cu}{\rm O}_2$ planes will magnify the intensity of a specific component of $\vec{Q}$ in X-ray scattering experiments \cite{PhysRevB.92.224504}. 

Within this formalism, we can define a composite SU(2) order parameter, $u_{SU(2)}=u\Delta_{SU(2)}$\cite{Efetov13}, where $u$ is:
\begin{equation}
u= \left(\begin{array}{cc} \phi & \psi\\ -\psi^* & \phi^* \end{array}\right)
\label{eq:COMop}
\end{equation}
The matrix $u$ is parametrized by two complex order parameters: the d-wave SC order parameter ($\psi$) and the d-wave CO order parameter ($\phi$). $u$ is a unitary matrix imposing a strong constraint on each of its components:
\begin{equation}
\phi^2+\psi^2=1
\label{eq:constraint1} 
\end{equation}
Thus, $u_{SU(2)}^2=\Delta_{SU(2)}^2$. The composite order parameter can be thought of as a pseudo-spin in four dimensions with two SC components and two CO components. $\Delta_{SU(2)}^2$ sets the length of this pseudo-spin. The length of this pseudo-spin can be described by a Ginzburg-Landau mean field theory. It goes to zero at a high mean field temperature, which we characterize as the pseudogap temperature ($T^*$)\cite{Efetov13}. $T^*$ controls the high energy physics of the problem. Below $T^*$, Eq.~\ref{eq:constraint1} describes a three dimensional hypersphere $\mathbb{S}^3$ in a four dimensional space. The transverse fluctuations of the composite order parameter on this hypersphere are described by an $O(4)$ non linear sigma model (NLSM) \cite{Efetov13}:
\begin{equation}
\frac{F}{T}=\frac{1}{t_0} \int tr[\nabla u^{\dagger} \nabla u + \kappa_{0} \tau_3 u^{\dagger} \tau_3 u] dR
\label{eq:NLSM}
\end{equation}
where $\kappa_{0}=(\alpha'_{\phi}-\alpha'_{\psi})/2$ is the difference of the zero temperature masses of the SC and CO fields, $t_{0}=2T/\rho_s^{0}$ is the scaled temperature, $\rho_s^{0}$ being the stiffness associated with spatial variation of the composite order parameter $u$, $\tau_3$ is the third Pauli spin matrix in the space of the matrix $u$, $tr$ is the trace over the space of $u$ and the integration is over the two dimensional real space coordinates. $\rho_s^{0}$ is proportional to $T^*$ \cite{Efetov13}. The free energy functional in Eq.~\eqref{eq:NLSM} has two primary contributions: 
\begin{itemize}

\item The first term  $tr[\nabla u^{\dagger} \nabla u]$ can be written in terms of the fields as $2(|\nabla \psi|^2+|\nabla \phi|^2)$. This term describes the spatial fluctuations of $\psi$ and $\phi$. If the mass of the SC field ($\alpha'_{\psi}$) is same as the mass of the CO field ($\alpha'_{\phi}$), i.e., $\kappa_{0}=0$, the SC and CO ground states are energetically degenerate resulting in an exact SU(2) symmetry. There is then no energy cost associated with the rotation of the pseudo-spin in the four dimensional space of the composite order parameter $u$. With $\kappa_{0}=0$, the two dimensional NLSM in Eq.~\eqref{eq:NLSM} at finite $t$ produces divergent fluctuations \cite{chaikin2000principles,Efetov13} destroying any long-range order in $\psi$ or $\phi$.

\item The second term $tr[\kappa_{0} \tau_3 u^{\dagger} \tau_3 u]$ can be written in terms of the fields as $2 \kappa_{0}(|\phi|^2-|\psi|^2)$. This term breaks the degeneracy between the SC and CO ground states. If $\kappa_0>0$, the pseudo-spin prefers the easy plane in the SC space characterized by a gapless Goldstone mode. If $\kappa_0<0$, the pseudo-spin prefers the easy plane in the CO space characterized by another gapless Goldstone mode. $\kappa_0$ introduces an anisotropy between the SC and CO easy planes. Thus, $\kappa_0$ defines the energy cost to rotate the pseudo-spin from one easy plane to the other and introduces a gap in the excitations of the pseudo-spin. This gap is small compared to the pseudogap energy scale ($T^{*}$) and the fluctuations governed by the first term in Eq.~\eqref{eq:NLSM} are still important indicating an approximate SU(2) symmetry. Since this anisotropy term in Eq.~\eqref{eq:NLSM} is quadratic in fields, we refer to its effect as {\it quadratic} symmetry breaking.

\end{itemize}

The energy difference between the two ground states can be further enhanced if the exact SU(2) symmetry is broken by the {\it biquadratic} terms in the free energy of the composite order parameter. The contribution from the biquadratic symmetry breaking in the free energy is given by:
\begin{equation}
\frac{F_{bq}}{T}=\frac{1}{t_0} \int z_{0} \left \{ \left ( tr[\tau_3 u^{\dagger} \tau_3 u] \right )^2 - 1 \right \} dR
\label{eq:biquadfr}
\end{equation}
where $z_{0}=(\beta-\gamma)/4$ with $\gamma$ being the coupling strength between the two orders and $\beta$ being the strength of the self interaction of both the fields. Expressing $u$ in terms of $\psi$ and $\phi$, Eq.~\eqref{eq:biquadfr} is given as $-4z_{0}|\psi|^2|\phi|^2$. If $\gamma=\beta$, $F_{bq}=0$ and the biquadratic terms do not contribute to the free energy. For $\gamma<\beta$, the gap in the excitations of the pseudo-spin is modified by the strength of the biquadratic symmetry breaking ($z_{0}$). In the parameter regime $-z_{0}<\kappa_{0}<z_{0}$, the total free energy ($F+F_{bq}$) accommodates a coexisting phase with both the SC and the CO being stable. The pseudo-spin prefers an intermediate direction making a finite angle with both the SC easy plane and the CO easy plane. On the other hand, if $\gamma>\beta$, the repulsion between the fields is large and there exists no coexistence and the situation is similar to the case when $F_{bq}=0$. We will assume that $z_{0}$ is small such that the approximate SU(2) symmetry is still valid for $T<T^*$.     
 
\subsection{Renormalization group treatment of the classical NLSM}\label{sec:CRG}

As discussed Sec.~\ref{sec:GL4}, the thermal fluctuations play a significant role in deciding the critical phenomenon in two spatial dimensions. We perform a renormalization group calculation to take care of these critical fluctuations described by the NLSM. In this section, we will not consider any time-dependent fluctuations nor the fluctuations in the modulus of the order parameters. Although, we will discuss the effects of time-dependent fluctuations in Sec.~\ref{sec:QRG}. Here, we will look at two cases of weak SU(2) symmetry breaking: a) only quadratic symmetry breaking ($\kappa_0 \ne 0$ and $z_{0}=0$), where the free energy will be given by Eq.~\eqref{eq:NLSM} b) both quadratic and biquadratic symmetry breaking ($\kappa_0 \ne 0$ and $z_{0} \ne 0$), where the total free energy is given by $F+F_{bq}$ ($F$ obtained from Eq.~\eqref{eq:NLSM} and $F_{bq}$ obtained from Eq.~\eqref{eq:biquadfr}).

First, we consider the case with only quadratic symmetry breaking ($z_{0}=0$). We treat the fluctuations around the mean field phase of the classical NLSM in Eq.~\eqref{eq:NLSM} using the renormalization group approach. We integrate out the fast varying components of the free energy in Eq.~\eqref{eq:NLSM} and write an effective slow varying counterpart with effective coupling constant $t$ and anisotropy parameter $\kappa$. Within one loop approximation, the RG flow equations (for details see appendix \ref{sec:RGcNLSM}) for the effective parameters are given by: 
\begin{equation}
\frac{dt}{dl}= \frac{t^2}{2 \pi} 
\label{eq:RG12d}
\end{equation}
\begin{equation}
\frac{d\left (ln\left (\frac{\kappa}{t}\right ) \right )}{dl}=-\frac{t}{ \pi} + 2
\label{eq:RG22d}
\end{equation}
where $l$ is the running logarithm variable of the RG. The solutions of Eqs.~\eqref{eq:RG12d} and \eqref{eq:RG22d} determine the flow of the renormalized parameters of the free energy. At $l=0$, $t=t_0$ and $\kappa=\kappa_0$, where $t_0$ and $\kappa_0$ are the bare values of the parameters. There is an ultraviolet momentum cutoff, $\Lambda$ which corresponds to the inverse of the minimum length of the theory. Additionally, there is an infrared cutoff $E_g^{1/2}$ where $E_g$ corresponds to the gap in the excitation spectrum. The RG flow of Eqs.~\eqref{eq:RG12d} and \eqref{eq:RG22d} stops at $l=ln(\Lambda/E_g^{1/2})$. The solutions of the effective parameters are:
\begin{equation}
t=t_0 \left ( 1- \frac{t_0}{2 \pi} ln\left ( \frac{\Lambda}{E_g^{1/2}} \right ) \right )^{-1}
\label{eq:Solutiont}
\end{equation}

\begin{equation}
\kappa=\kappa_0 \left ( \frac{\Lambda}{E_g^{1/2}} \right )^2 \left ( 1- \frac{t_0}{2 \pi} ln\left ( \frac{\Lambda}{E_g^{1/2}} \right ) \right )
\label{eq:Solutionm}
\end{equation}
The divergence of the effective coupling constant $t$ in Eq. \eqref{eq:Solutiont} can be seen as an evidence of a transition from an ordered phase to a disordered phase. Along with the divergence of $t$, the effective anisotropy $\kappa$ also goes to zero. The system, thus goes to a mixture of fluctuating SC and CO with no long-range order, which is characterized as the pseudogap phase.

If $\kappa_0>0$, only the SC phase is stable. In this phase, we study the fluctuations around the corresponding mean field solution of $u$. In the absence of external magnetic field, the gap in the excitations of the pseudo-spin corresponds to the difference of the masses of the SC and CO fields. So, $E_g=2 \kappa_0$. The transition temperature from the SC phase to the pseudogap phase i.e., the temperature where the effective coupling constant $t$ diverges is given by:
\begin{equation}
T_{c}=\frac{2 \pi \rho_s^0}{ln \left ( \frac{\Lambda^2}{2\kappa_{0}} \right )}
\label{eq:tr1}
\end{equation}
At this temperature, the anisotropy $\kappa$ also goes to zero.  The pseudogap temperature ($T^*$) is controlled by $\rho_s^0$ and thus can be significantly higher than $T_{c}$.
 
In the presence of an external magnetic field, the gap in the excitation spectrum $E_g$ or the energy required to break the long-range SC coherence is replaced by:
\begin{equation}
E_g^{sc}=2 \kappa_0-\zeta B
\label{eq:Eg1}
\end{equation}
where $\zeta$ is a constant. In the presence of magnetic field, the SC order parameter becomes inhomogeneous below a length scale which is given by the coherence length ($\xi$) of the superconductor. Hence, the minimum length of this effective homogeneous RG analysis is constrained by $\xi$ and thus the upper momentum cutoff is given by $\Lambda=\xi^{-1}$. In a cuprate superconductor, $\xi$ is quite small compared to the penetration depth. The transition magnetic field ($B_{sc}$) where $t$ diverges is given by:
\begin{equation}
B_{sc}=B^{0} \left \{ 1 - \frac{1}{2 \xi^2 \kappa_0} exp \left ( -\frac{2 \pi \rho_s^0}{T} \right ) \right \}
\label{eq:tr2}
\end{equation} 
where the SC phase is stabilized for $B<B^{0}$ and $B^{0}=(2\kappa_0)/\zeta$. 
If the ground state is the CO phase, we have to consider the fluctuations of the nonlinear sigma model around the mean field solution $u=\mathbb{I}$. RG equations and solutions are equivalent to the case when the ground state is the SC phase, but $E_g$ is now replaced by:
\begin{equation}
E_g^{co}=-2 \kappa_0+\zeta B
\label{eq:Eg2}
\end{equation}
The transition magnetic field ($B_{co}$) is thus given by:
\begin{equation}
B_{co}=B^{0} \left \{ 1 + \frac{1}{2 \xi^2 \kappa_0} exp \left ( -\frac{2 \pi \rho_s^0}{T} \right ) \right \}
\label{eq:trea3}
\end{equation} 
where the CO phase is stabilized for $B>B^{0}$.

\begin{figure}[h]
\includegraphics[width=0.4\textwidth]{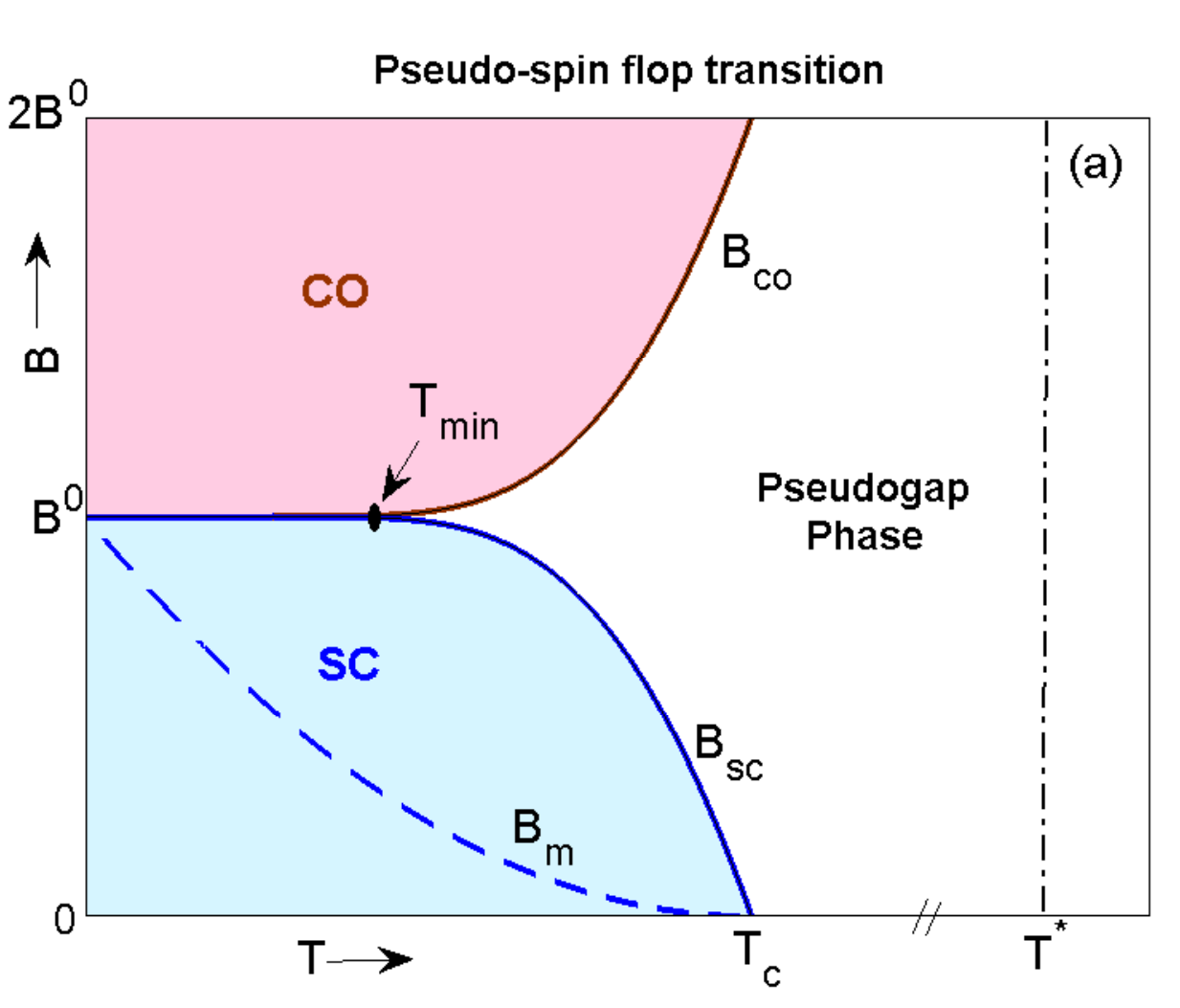} \\
\includegraphics[width=0.4\textwidth]{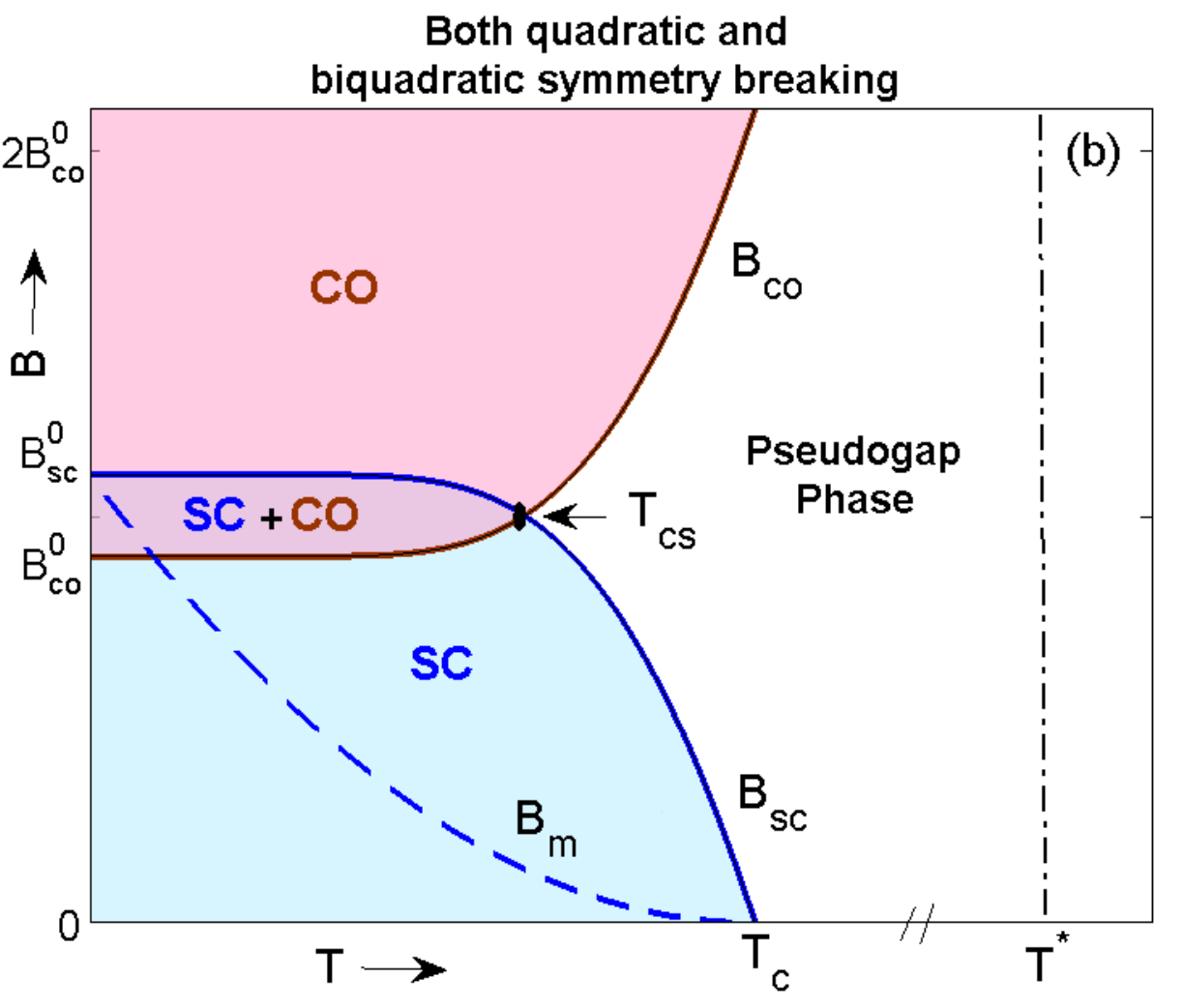}
\caption{ The B-T phase diagram obtained within a renormalization group treatment of the classical nonlinear sigma model. (a): Anisotropy ($\kappa_0$) between the masses of the SC and CO fields induces a quadratic symmetry breaking at $T=0$ and $B=0$. Increasing the magnetic field destroys the SC order giving rise to the CO marked by a pseudo-spin flop transition at $B=B^0$ for $T<T_{min}$. $B_{co}$ remain flat at low $T$ due to suppressed thermal fluctuations and rises steeply for $T>T_{min}$. The thermal fluctuations drive the anisotropy to zero on the $B_{co}$ and $B_{sc}$ lines. As a result, the system hesitates between the CO phase and the SC phase with no visible long-range order marking the pseudogap phase with SU(2) fluctuations for $T<T^*$. (b): If the coupling strength ($\gamma$) between the SC and CO is not exactly equal to the coefficient ($\beta$) of each biquadratic terms, the biquadratic SU(2) symmetry is also broken ($z_0 \ne 0$). $B_{sc}^0$ and $B_{co}^0$ (the transition fields at $T=0$) are different with a region of coexistence in between for $T<T_{cs}$ for $\gamma<\beta$. The renormalized effective anisotropy between the CO and the SC fields become zero at $T=T_{cs}$ and the SU(2) fluctuations are observable for $T_{cs}<T<T^{*}$.
If $\gamma>\beta$, the strong repulsion between the fields destabilizes any coexisting phase with a pseudo-spin flop transition at $B=B^0$ and the B-T phase diagram is exactly same as in (a). We sketch the phenomenological temperature dependence of the vortex melting transition field $B_m$ to distinguish the upper critical field $B_{sc}$ from the melting transition, see text for details.
}
\label{fig:crg}
\end{figure}

It is important to note that $B_{co}(T=0)=B_{sc}(T=0)=B^{0}$. $B^0$ is the zero temperature upper critical field for the superconductor in the presence of strong competition with the CO phase. At $T=0$, the ground state of the system is the SC phase for $B<B^0$ and the ground state is the CO phase for $B>B^0$. Thus, there is no coexisting phase in the B-T phase diagram if only the quadratic symmetry is broken ($\kappa_0 \ne 0,z_0=0$). In terms of the composite order parameter, the pseudo-spin flops from a direction aligned in the SC easy plane to a direction aligned in the CO easy plane at $B=B^0$. The thermal fluctuations are absent at $T=0$. Thus, we expect the mean field solutions for the transition fields should give the same result as the solutions obtained in Eqs.~\eqref{eq:tr2} and~\eqref{eq:trea3} at $T=0$. The value of the transition field, $B^0$, is found to be same within a mean field picture if we use the constraint $\psi^2+\phi^2=1$ with weak quadratic symmetry breaking ($\kappa_0$), as shown in Eq.~\eqref{eq:conb0} of the appendix \ref{sec:BCwithcon}. In this section, we choose the same range of the masses $\alpha'_{\psi}$ and $\alpha'_{\phi}$ as in Sec.~\ref{sec:GL2} to have a correct comparison with the GL theory.    

We now discuss the temperature dependences of $B_{sc}$ in Eq.~\eqref{eq:tr2} and $B_{co}$ in Eq.~\eqref{eq:trea3}. First, we note that $B_{sc}$ falls and $B_{co}$ increases with increase in temperature. The rate of change of $B_{sc}$ and $B_{co}$ with $T$ is exactly the same as the two orders are strongly constrained by the SU(2) symmetry. The corrections to the $T=0$ value of $B_{co}$ vanish exponentially fast as $T \rightarrow 0$. Therefore, $B_{co}$ (and $B_{sc}$) does not differ from $B^0$ and remains flat for small $T$ ($T\ll\rho_s^0$). In comparison to this unique result obtained in SU(2) theory, the flatness is specific to fined tuned mass parameters in a GL mean field theory (Sec.~\ref{sec:GL2}), which ignores the thermal phase fluctuations. 

If $T \sim \rho_s^0$, the exponential terms in Eqs.~\eqref{eq:tr2} and~\eqref{eq:trea3} are no longer small. In this regime of temperatures, the increase of $B_{co}$ is regulated by the coefficient $(2\kappa_0 \xi^2)^{-1}$ of the exponential term. Since both $\kappa_0$ and $\xi$ are small, there is a steep rise in $B_{co}$ for $T \sim \rho_s^0$. As a result, at high magnetic fields, the temperature corresponding to the transition from the CO phase to the pseudogap phase is fairly insensitive to the magnetic fields. We name this transition temperature as $T_{co}$ where $B_{co}=2B^{0}$. We obtain $T_{co}$ using Eq.~\eqref{eq:trea3} and compare it with $T_{c}$ obtained from Eq.~\ref{eq:tr1}: 
\begin{equation}
T_{co}=\frac{2 \pi \rho_s^0}{ln \left ( \frac{1}{2\xi^2\kappa_0} \right )}=T_c
\label{eq:tch1}
\end{equation}
$T_{co}$ and $T_{c}$ are the temperatures where the thermal fluctuations renormalize the difference of the masses (and not the individual masses) to zero. So, within this SU(2) formalism, we expect $T_{co}=T_{c}$ (obtained in Eq.\eqref{eq:tch1}) as the orders are destroyed by the thermal phase fluctuations. In contrast, $T_{co}$ and $T_{c}$ in a GL theory are the temperatures where the amplitudes of the CO order parameter and the SC order parameter go to zero respectively. The vanishing amplitudes are reflected by the vanishing masses ($\alpha_{\psi}$ and $\alpha_{\phi}$) and are not connected to each other. 

We will now analyze the case where both quadratic ($\kappa_0 \ne 0$) and biquadratic ($z_{0} \ne 0$) symmetries are broken. For the simplicity of our calculation, we will make two assumptions. First, the parameter $z_{0}$ does not get renormalized during the RG process. Second, we will ignore the effect of $z_0$ on the renormalization of $\kappa_0$. We will only consider the renormalization of the parameters $t_0$ and $\kappa_0$. The RG flow equations will remain the same as in Eqs.~\eqref{eq:RG12d} and~\eqref{eq:RG22d}. But the presence of $F_{bq}$ (Eq.~\eqref{eq:biquadfr}) in the total free energy affects the gap in the excitation spectrum $E_g$. Due to the constraint in the fields, the mass terms corresponding to each field are modified by $z_0$ (also see the appendix \ref{sec:BCwithcon}). If $z_0<0$ (recalling that $4z_0=\beta-\gamma$), the repulsion between the fields is large. As a result, the zero temperature gap in the excitation spectrum remains similar to the case of $z_0=0$. The transition fields $B_{sc}$ and $B_{co}$ are governed by Eqs.~\eqref{eq:tr2} and~\eqref{eq:trea3} with no region of coexistence. If $z_0>0$, the gaps in the excitation spectrum will be given by:
\begin{equation}
E_g^{sc}=2 \kappa_0 +4 z_0 -\zeta B, E_g^{co}=-2 \kappa_0+4 z_0+\zeta B 
\label{eq:Eg3}
\end{equation}
and the corresponding transition magnetic fields are given by:        
\begin{equation}
B_{sc}=B_{sc}^{0} \left \{ 1 - \frac{1}{ \xi^2 (2\kappa_0+4 z_0)} exp \left ( -\frac{2 \pi \rho_s^0}{T} \right ) \right \}
\label{eq:tr3}
\end{equation}
and
\begin{equation}
B_{co}=B_{co}^{0} \left \{ 1 + \frac{1}{ \xi^2 (2\kappa_0-4 z_0)} exp \left ( -\frac{2 \pi \rho_s^0}{T} \right ) \right \}
\label{eq:tr4}
\end{equation} 
where $B_{sc}^0=(2\kappa_0+4z_0)/\zeta$ and $B_{co}^0=(2\kappa_0-4z_0)/\zeta$.

Since $B_{sc}(T=0) < B_{co}(T=0)$, there is a coexisting phase in a short range of magnetic fields for low $T$. $B_{sc}^0$ and $B_{co}^0$ define the zero temperature magnetic fields corresponding to the transitions from the SC phase to the coexisting phase and the coexisting phase to the CO phase respectively. The values of $B_{sc}^0$ and $B_{co}^0$ obtained here match their corresponding values derived in a mean field treatment using the constraint $\psi^2+\phi^2=1$ (see Eq.~\eqref{eq:bscconco} and Eq.~\eqref{eq:bcoconco} of the appendix \ref{sec:BCwithcon}). From Eqs.~\eqref{eq:tr3} and~\eqref{eq:tr4}, $T_c$ (where $B_{sc}=0$) and $T_{co}$ (where $B_{co}=2B_{co}^0$) are given by:    
\begin{equation}
T_{c}=\frac{2 \pi \rho_s^0}{ln \left ( \xi^{-2}(2\kappa_0+4z_0)^{-1} \right )}, T_{co}=\frac{2 \pi \rho_s^0}{ln \left ( \xi^{-2}(2\kappa_0-4z_0)^{-1} \right )}
\label{eq:tch2}
\end{equation}
As $z_0>0$, $T_{co}<T_{c}$. Motivated by experimental facts \cite{2014NatCo...5E3280G,Wu13a}, we assume that the region of coexistence ($B_{sc}^0-B_{co}^0$) is small compared to the upper critical field, $B_{sc}^0$. As a result, $z_0 \ll \kappa_0$ which implies $T_{co} \approx T_{c}$.

\subsubsection*{\textbf{\textit{B-T phase diagram: flatness, coexistence and SU(2) symmetry}}}

We use the expressions for the transition magnetic fields, $B_{sc}$ and $B_{co}$, obtained within the RG analysis of the classical NLSM to construct the B-T phase diagram. 

In Fig.~\ref{fig:crg}(a), we show the B-T phase diagram for $z_0=0$ (when only quadratic symmetry is broken). We first discuss the transition from the SC phase to the CO phase at $T=0$. The ground state of the NLSM in Eq.~\eqref{eq:NLSM} is determined by the value of the anisotropy ($\kappa_0$) between the masses of the SC and CO fields. An external magnetic field effectively renormalizes the mass of the SC field and in turn renormalizes $\kappa_{0}$ to $\kappa_{0}^{\rm eff}=\kappa_{0}-\zeta B/2$. At low $B$, $\kappa_{0}^{\rm eff}>0$. As a result, the ground state is the SC phase. $\kappa_{0}^{\rm eff}$ decreases with increasing $B$ and reaches $\kappa_{0}^{\rm eff}=0$ at $B=B^0$. For $B>B^0$, $\kappa_{0}^{\rm eff}$ becomes negative and favors the CO phase as the ground state. The pseudo-spin flops from a direction aligned in the SC easy plane to a direction aligned in the CO easy plane at $B=B^0$.

If the system favors one phase as the ground state, the other phase remains as a metastable state with a higher energy compared to the ground state. If we increase $T$, the probability of the system seeing this metastable state increases due to the thermal fluctuations. Within the RG technique described above, these thermal fluctuations are captured by renormalizing $\kappa_{0}^{\rm eff}$ to $\kappa^{\rm eff}$ (note that $B$ is an external parameter and is not renormalized). If $\kappa^{\rm eff}=0$, the SC phase and the CO phase become degenerate in energy, recovering the exact SU(2) symmetry between the SC and the CO. This marks the transition to a disordered phase where the system hesitates between the SC and the CO. We characterize this disordered phase as the pseudogap phase which has no visible long-range order (shown as white region below $T=T^*$ in Fig.~\ref{fig:crg}). $B_{sc}$ and $B_{co}$ in Eqs.~\eqref{eq:tr2} and~\eqref{eq:trea3} determine the transitions from the SC phase and the CO phase to the pseudogap phase respectively. At low $T$, the thermal fluctuations are weak and $\kappa^{\rm eff}$ remains insensitive to temperature. Consequently, $B_{sc}$ and $B_{co}$ remain flat up to a temperature $T_{\rm min}$ ($\propto \rho_s^0$). For $T>T_{\rm min}$, due to strong thermal fluctuations, $\kappa^{\rm eff}$ gets renormalized strongly and the line $B_{co}$ rises (or $B_{sc}$ falls) steeply. 
 
We now discuss the B-T phase diagram for $z_0 \ne 0$ (when both quadratic and biquadratic symmetries are broken), as show in Fig.~\ref{fig:crg}(b). If $z_0<0$, strong repulsion between the fields does not favor any phase with coexisting SC and CO. In this case, the phase diagram remains the same as in Fig.~\ref{fig:crg}(a) with a pseudo-spin flop transition at $B=B^0$. 

The possibility of the presence of a coexisting phase emerges with $z_0>0$. At $T=0$, $B^0$ is split to $B_{co}^0$ and $B_{sc}^0$ corresponding to $\kappa_{0}^{\rm eff}=-z_0$ and $\kappa_{0}^{\rm eff}=z_0$ respectively. For $B_{co}^0<B<B_{sc}^0$, the ground state is a coexistence phase with both the SC order and the CO. The pseudo-spin orients in the SC plane for $B<B_{co}^0$, changes its orientation to a direction making an angle with both the SC and CO planes for $B_{co}^0<B<B_{sc}^0$ and then finally orients itself in the CO plane for $B>B_{sc}^0$. The two transition lines $B_{co}$ and $B_{sc}$ intersect at a temperature $T_{cs}$ (shown in Fig.~\ref{fig:crg}(b)) where $\kappa^{\rm eff}=0$ with an exact SU(2) symmetry. The system goes into the pseudogap phase for higher temperatures ($T_{cs}<T<T^*$).

Interestingly, the region of coexistence for $T<T_{cs}$ has both broken SU(2) symmetry and broken U(1) symmetries. U(1) symmetry breaking corresponding to the CO field will result into a phase with a diagonal long-range order and such symmetry breaking for the SC field will favor superconductivity with an off-diagonal long-range order. This phase shows supersolidity and will also show superconducting properties like zero resistance \cite{0953-8984-8-26-012}. Similar coexistence is also found in an attractive Hubbard model \cite{PhysRevLett.66.3203,PhysRevLett.71.4238} where at half filling there is an exact SU(2) symmetry. 

The local signatures of this coexisting phase can be observed in the halo regions surrounding each vortex core. The superconducting order parameter is expected to show periodic modulations (commonly known as pair density waves \cite{2016Natur.532..343H,Wang15a,Wang15b}) in the vortex halo \cite{2018arXiv180204673E}. The wave vector corresponding to this modulation is connected to the charge modulation wave vector. This coexisting phase also has unique signatures in the measurements of the collective modes, with observation of two massless phasons (or Goldstone Bosons) and two massive amplitudons (or pseudo-Goldstone Bosons). A detailed study of these collective modes will be reported elsewhere.

The width of the flatness of $B_{co}$ and $B_{sc}$ is proportional to the stiffness $\rho_s^0$ in the NLSM. In this paper, we derive the transition lines using the same $\rho_s^0$ for the SC order and the CO order. In contrast, if $\rho_s^0$ for the SC field is taken smaller than the $\rho_s^0$ for the CO field, $B_{co}$ will remain flat for a larger temperature window compared to $B_{sc}$.

\subsubsection*{\textbf{\textit{Vortex melting}}}

In Fig.~\ref{fig:crg}, $B_{sc}$ is the upper critical field, below which the pairing gap remains finite and vortices start to appear in the system. If these vortices arrange themselves to form a lattice, the electric resistance goes to zero. Interestingly, the vortex lattice melts to form a vortex liquid due to the thermal fluctuations at a magnetic field $B_{m}$. In cuprates, the two fields $B_{sc}$ and $B_m$ are not the same and there exists a region in the B-T phase diagram where the electric resistance is not zero but the pairing gap is finite. While the vortex melting field \cite{RevModPhys.82.109,RevModPhys.66.1125} is easily observed in transport experiments \cite{PhysRevLett.76.835,PhysRevLett.69.824,PhysRevB.86.174501}, the direct detection of the upper critical field has been challenging until recently when it was detected from the thermal conductivity measurements \cite{2014NatCo...5E3280G}. In our RG treatment, we do not consider the thermal melting transition temperature. Instead, we give an idea of the upper critical field only. In the following paragraph, we present a phenomenological way of sketching the $B_{m}$(T) line in the B-T phase diagram. 

In a bulk three dimensional sample, the vortices form flux lines (where the magnetic field penetrates the sample) aligned along the direction of the applied magnetic field. The position of these lines will vibrate about their mean position of the vortex lattice due to thermal fluctuations. We will use Lindemann criterion \cite{LindCrit}, where melting is characterized by equating the amplitude of this vibration to a considerable fraction of the spacing between the vortices \cite{book:17888}. The vortex lines repel each other and once there is a distortion from the equilibrium lattice positions, they will experience a restoring force ($\Sigma$) per unit length. Also, there is an energy cost to bend a segment ($l$) of these vortex lines, given in terms of the line tension $\Omega$. So, the total energy cost to displace a segment of vortex line by a small displacement $\delta$ is $\Sigma \delta^2 l + \Omega \delta^2/l$. The optimum length of segment displaced, determined by minimizing this total energy cost, is $l^2=\Omega/\Sigma$. This optimal energy cost has to be equal to the thermal energy $k_{B}T$. Using this equality, we get the displacement due to thermal vibrations as $\delta^2 \sim k_{B}T/(\Sigma \Omega)^{1/2}$. We will use estimates of the line tension and the restoring force from the conventional GL phenomenology. They are given as $\Sigma \sim B/(\lambda^2)$ and $\Omega \sim 1/(\lambda^2)$, where $\lambda$ is the penetration depth of the superconductor. Now, the magnetic flux per unit cell of the vortex lattice is a universal constant, so the separation between vortices are given as $d_{av} \sim 1/(B^{1/2})$. Using Lindemann criterion, the displacement of the flux lines should be proportional to the separation between the vortices, $\delta \sim d_{av}$. This gives an estimate of the melting transition magnetic field, $B_{m} \sim \lambda^{-4}T^{-2}$. Near $T_c$, $\lambda^{-2} \sim (T_c-T)$. So, near $T_c$, $B_{m} \sim (T_c-T)^2$. On the other hand, within a GL theory, the upper critical field, $B_{c2} \sim \lambda^{-2} \sim (T_c-T)$ and thus $B_m<B_{c2}$ for $T<T_c$. In Fig.~\ref{fig:crg}, we plot the melting transition line, $B_{m}=B_{sc}^{0}(1-T/T_c)^2$. At $T=0$, $B_m=B_{sc}^0$, which is motivated by the experimental findings of Ref.~\onlinecite{2014NatCo...5E3280G}.    

Throughout the analysis in the preceding paragraph, we have assumed the system to be a 3D bulk sample. In practice, the anisotropy \cite{RevModPhys.66.1125} between the directions along the 2D ${\rm Cu}{\rm O}_2$ planes and the direction perpendicular to them also plays a role in determining $B_m$. In a conventional superconductor along with some cuprates with lower anisotropy, the difference between $B_m$ and $B_{c2}$ is indistinguishable within the experimental accuracy close to $T_c$. If the anisotropy is large as in most of the cuprates, the difference is prominent and looks similar to our sketch in Fig.~\ref{fig:crg}. This simple analysis does not take into account the features like pinning and long-range nature of vortex-vortex interactions, which have been studied extensively in the literature \cite{RevModPhys.66.1125,RevModPhys.82.109}.

\subsection{Renormalization group treatment of the quantum NLSM}\label{sec:QRG}

In the RG treatment of the classical NLSM in Sec.~\ref{sec:CRG}, we considered the spatial fluctuations of the composite order parameter $u$ and neglected its temporal fluctuations. Here, we consider the quantum mechanical NLSM by treating the free energy \cite{Chakravarty89}:
\begin{equation}
F=\frac{\rho_s^0}{2} \int_0^{\beta} d\tau \int tr[ \nabla u^{\dagger} \nabla u + \frac{1}{c^2}\left | \frac{\partial u}{\partial \tau} \right |^2 + \kappa_0 \tau_3 u^{\dagger} \tau_3 u] dR
\label{eq:QNLSM1}
\end{equation} 
where $\beta=1/T$, $\tau$ is the imaginary time and we have used the units $\hbar=1$ and $k_{\text{B}}=1$. $c$ is the velocity of the fluctuation modes that defines the perpendicular susceptibility $\chi=\rho_s^0/c^2$. While $\rho_s^0$ defines the length scale of the fluctuations of $u$, $\chi$ defines the time scale of the corresponding fluctuations. By rescaling $R$ and $\tau$, such that they are of the same dimensions and the wave vector cut off is unity, we get:
\begin{equation}
F=\frac{1}{2g_0} \int_0^{v} d\tau \int tr[\nabla u^{\dagger} \nabla u + \left | \frac{\partial u}{\partial \tau} \right |^2 + \bar{\kappa}_{0} \tau_3 u^{\dagger} \tau_3 u] dR 
\label{eq:QNLSM2}
\end{equation} 
where $g_0=c\Lambda^{d-1}/\rho_s$, $v=c\beta \Lambda$, $\bar{\kappa}_0=\kappa_0 \Lambda^{-2}$. If $v \rightarrow 0$, the quantum fluctuations are not important, i.e., the configurations that contribute significantly to the partition function are independent of $\tau$ and $F$ maps to the classical NLSM in Eq.~\eqref{eq:NLSM} with coupling constant $2g_0/v$. On the other hand, $v \rightarrow 1$ means the quantum fluctuations can no longer be neglected.   

\begin{figure}[t]
\includegraphics[width=0.4\textwidth]{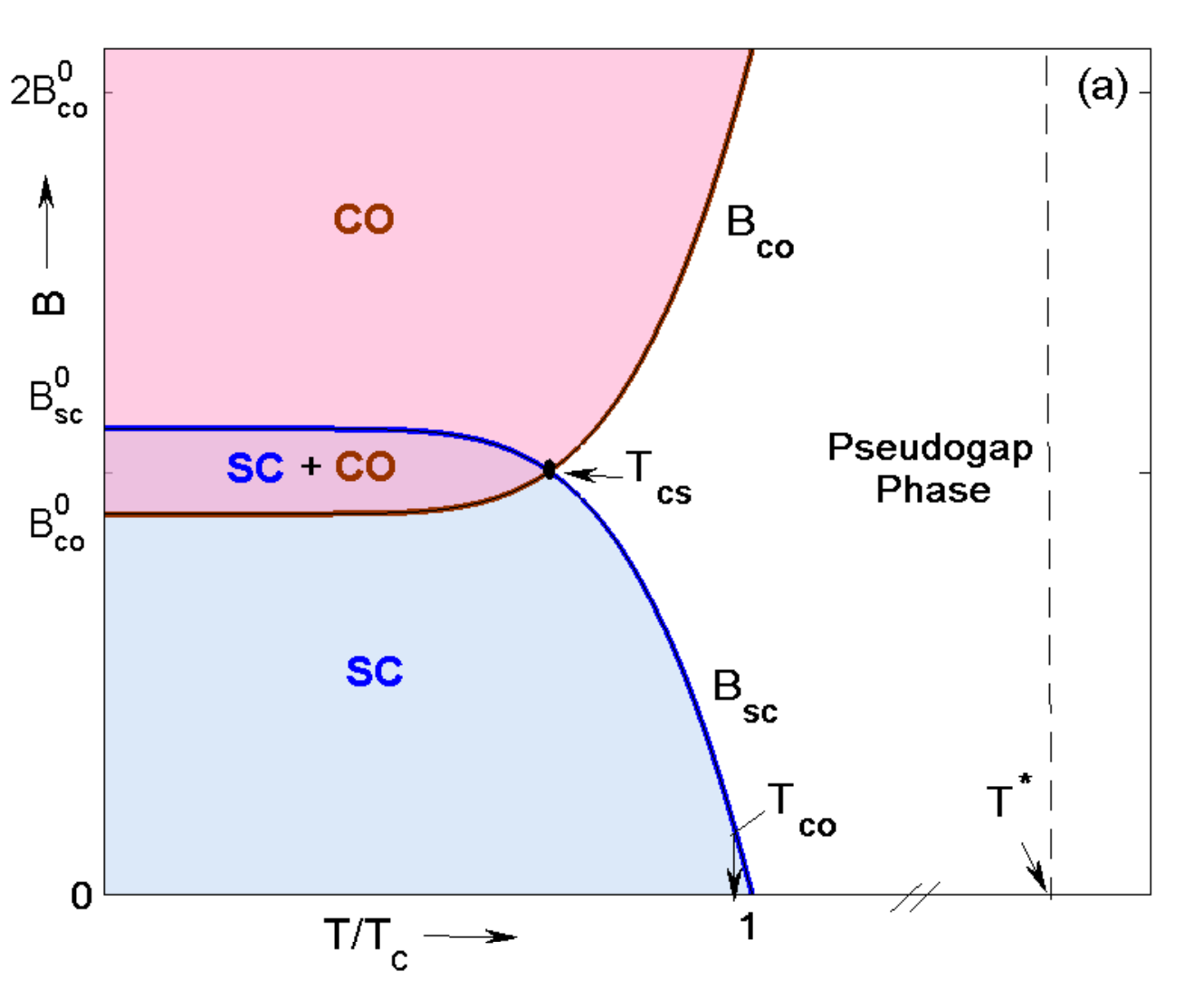} \\
\includegraphics[width=0.4\textwidth]{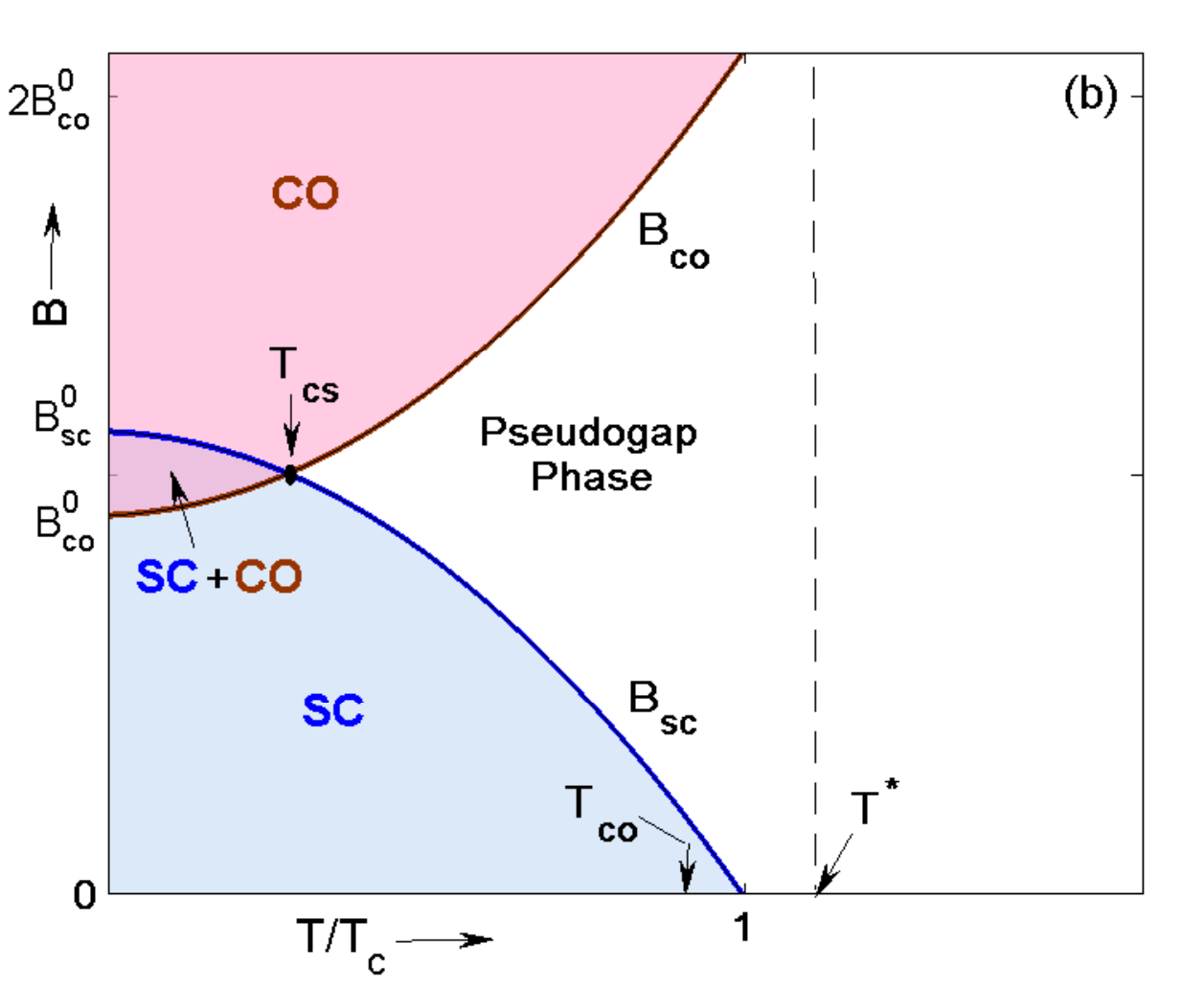}
\caption{ The B-T phase diagram obtained within a renormalization group treatment of the quantum mechanical nonlinear sigma model with both the quadratic ($\kappa \ne 0$) and the biquadratic symmetry breaking ($z_0 \ne 0$) for (a) $g_0/t_0 \rightarrow 0$ (extreme classical limit) and (b) $g_0=0.5 g_c$ (regime closer to the quantum critical point). The width of flatness in $B_{co}$ or $B_{sc}$ and the region of coexistence are both restricted to a smaller temperature window in (b). The temperature ($T_{cs}$), above which the SU(2) fluctuations become observable, decreases as we increase the strength of the quantum fluctuations. The CO transition temperature at high magnetic fields becomes sensitive to the corresponding magnetic field in (b) and no longer remains independent of $B$ as in (a). The CO transition temperature ($T_{co}$), at $B=2B_{co}^0$, becomes smaller than $T_c$ as we approach the quantum critical point. Note that we have a break in the temperature axis in (a) for $1<(T/T_c)<T^*$ and no such break in (b). This indicates that, in the extreme classical limit, $T^* \gg T_c$, but the ratio $T^*/T_c$ reduces significantly as one approaches the quantum critical point.
}
\label{fig:qrg}
\end{figure}

We use the same renormalization treatment of integrating out fast variables as in Sec.~\ref{sec:CRG} within a time slab thickness of $v$. We impose a upper momentum cutoff $\Lambda$ as in Sec.~\ref{sec:CRG}, but do not impose any cutoff on the time scale as the quantum fluctuations exist at all time scales. The RG differential equations (see the appendix \ref{sec:RGqNLSM}) are given in terms of a dimensionless parameter $t=T \Lambda^{d-2}/\rho_s=g/v$ by:
\begin{equation}
\frac{dv}{dl}=-v
\label{eq:QRG1}
\end{equation} 
\begin{equation}
\frac{dt}{dl}= \frac{1}{4 \pi} v t^2 coth \left ( \frac{v}{2} \right )
\label{eq:QRG2}
\end{equation} 
\begin{equation}
\frac{d\left (ln\left (\frac{\bar{\kappa}}{t}\right ) \right )}{dl}=-\frac{t}{ 2 \pi} v coth \left ( \frac{v}{2} \right )  + 2
\label{eq:QRG3}
\end{equation} 
The thickness of the time slab ($v$) renormalizes trivially by a factor of $b$ ($v=v_0 b^{-1}$), where $v_0=c \Lambda/T$. The classical limit of Eq.~\eqref{eq:QRG2} is obtained by taking $v=g/t \rightarrow 0$. In this limit, Eq.~\eqref{eq:QRG2} maps to Eq.~\eqref{eq:RG12d}. Using Eqs.~\eqref{eq:QRG1} and \eqref{eq:QRG2}, the RG equation for the coupling constant $g$ in two spatial dimensions at zero temperature ($T=0$) can be obtained as:
\begin{equation}
\frac{dg}{dl}= - g + \frac{g^2}{2 \pi} 
\label{eq:QRG4}
\end{equation} 
Eq.~\eqref{eq:QRG4} has a non-trivial fixed point at $g=g_c=2 \pi$. This point describes a quantum transition from a disordered phase to an ordered phase at zero temperature. 
At finite temperature, the RG flow of Eqs. \eqref{eq:QRG2} and \eqref{eq:QRG3} stops at $l=ln(\Lambda/E_g^{1/2})$, yielding the solutions:
\begin{equation}
t=t_0 \left ( 1+\frac{t_0}{2 \pi} ln \left | \frac{sinh\left ( \frac{v_0 E_g^{1/2}}{2 \Lambda} \right )}{sinh\left ( \frac{v_0}{2} \right )} \right | \right )^{-1}
\label{eq:SolQRG1}
\end{equation}

\begin{equation}
\kappa=\kappa_0 \left ( \frac{\Lambda}{E_g} \right )^2 \left ( 1+\frac{t_0}{2 \pi} ln \left | \frac{sinh\left ( \frac{v_0 E_g^{1/2}}{2 \Lambda} \right )}{sinh\left ( \frac{v_0}{2} \right )} \right | \right )
\label{eq:SolQRG2}
\end{equation}

We expect a thermal transition at finite temperature as obtained in Sec.~\ref{sec:CRG}.
We consider the case where both the quadratic ($\kappa_0 \ne 0$) and the biquadratic ($z_0 \ne 0$) symmetries are broken. From Eq.~\eqref{eq:Eg3}, the gap in the excitation spectrum in the presence of external magnetic field will be given by $E_g^{sc}=2 \kappa_0 +4 z_0 -\zeta B$ and $E_g^{co}=-2 \kappa_0+4 z_0+\zeta B$ in the SC and CO phases respectively. The transition magnetic fields are given by:
\begin{equation}
B_{sc}=B_{sc}^{0} \left [ 1-  \frac{4 T^2}{c^2(2\kappa_0 + 4 z_0)} R^2 \right ]
\label{eq:tr4q}
\end{equation}
and
\begin{equation}
B_{co}=B_{co}^{0} \left [ 1+  \frac{4 T^2}{c^2(2\kappa_0 - 4 z_0)} R^2 \right ]
\label{eq:tr5}
\end{equation}
where
\begin{equation}
R=sinh^{-1} \left \{ sinh \left ( \frac{c\Lambda}{2T} \right ) exp \left ( -\frac{2 \pi \rho_s^0}{T} \right ) \right \}
\label{eq:tr6}
\end{equation}
For only quadratic symmetry breaking ($z_0=0$) and in the case of $z_0<0$, $B_{sc}^0=B_{co}^0=B^0$. In the regime $g_0<g_c$ and $g_{0}/g_c<1-t_0/(2\pi)$, $R$ in Eq.~\eqref{eq:tr6} can be written as: 
\begin{equation}
R=exp \left ( -\frac{2\pi \tilde{\rho}_s}{T} \right )
\label{eq:tr7}
\end{equation}
and gives $B_{sc}$ and $B_{co}$ similar to Eq.~\eqref{eq:tr3} and \eqref{eq:tr4} but with upper momentum cutoff going as $T/c$ and the stiffness is renormalized due to quantum fluctuations at zero temperature as $\tilde{\rho_s}=\rho_s^0(1-g_0/g_c)$. This regime is often referred to as renormalized classical \cite{Chakravarty89} regime of interest. We set the values of $c$, $\Lambda$ and $\rho_s^0$ in such a way that we get the temperature dependence of $B_{sc}$ and $B_{co}$ in the extreme classical regime of $v_0 \rightarrow 0$ the same as in Fig.~\ref{fig:crg}(b). 

\subsubsection*{\textbf{\textit{B-T phase diagram: effect of quantum critical point}}}

In Fig.~\ref{fig:qrg}, we show the B-T phase diagram obtained from the quantum mechanical NLSM for $z_0 \ne 0$. We show the extreme classical limit in Fig.~\ref{fig:qrg}(a) and the case with $g_0$ close to $g_c$ in Fig.~\ref{fig:qrg}(b).

As we observed in Sec.~\ref{sec:CRG}, the transition magnetic fields $B_{sc}$ and $B_{co}$ remain independent of the temperature up to a temperature scale $T_{cs}$ in a classical NLSM. In a classical NLSM, the temperature scales as the stiffness $\rho_s$, which describes the length scale over which the mean field phase of NLSM fluctuates. 

We now discuss the effect of the pseudogap quantum critical point under the SC dome. In order to capture the effects of this quantum critical point, we have to incorporate also the time scale over which the mean field phase of NLSM fluctuates. At all non-zero temperatures, all the quantum fluctuations can be integrated out to obtain an effective classical NLSM. But this effective classical NLSM is described in terms of a renormalized stiffness $\tilde{\rho}_s$, which contains the effects of quantum fluctuations at zero temperature. At the pseudogap quantum critical point, $\tilde{\rho}_s \rightarrow 0$. As a result, the range of temperature ($T_{cs}$), over which $B_{sc}$ or $B_{co}$ remains flat, reduces along with the reduction in the coexistence region, as seen in Fig.~\ref{fig:qrg}(b). $B_{sc}$ or $B_{co}$ varies quadratically with temperature close to the critical point. As a consequence, the CO transition temperature $T_{co}$ at high magnetic field becomes sensitive to the magnetic field at which the measurement is performed. In Fig.~\ref{fig:qrg}, we characterize $T_{co}$ as the temperature at which $B_{co}=2B_{co}^0$. If we are in the extreme classical limit, $T_{co} \approx T_{c}$ as explained in Sec.~\ref{sec:CRG}. Once the effect of quantum critical point is dominant as in Fig.~\ref{fig:qrg}(b), $T_{co}$ becomes much smaller than $T_{c}$.   

\begin{figure}[t]
\includegraphics[width=7.4cm,height=6.0cm]{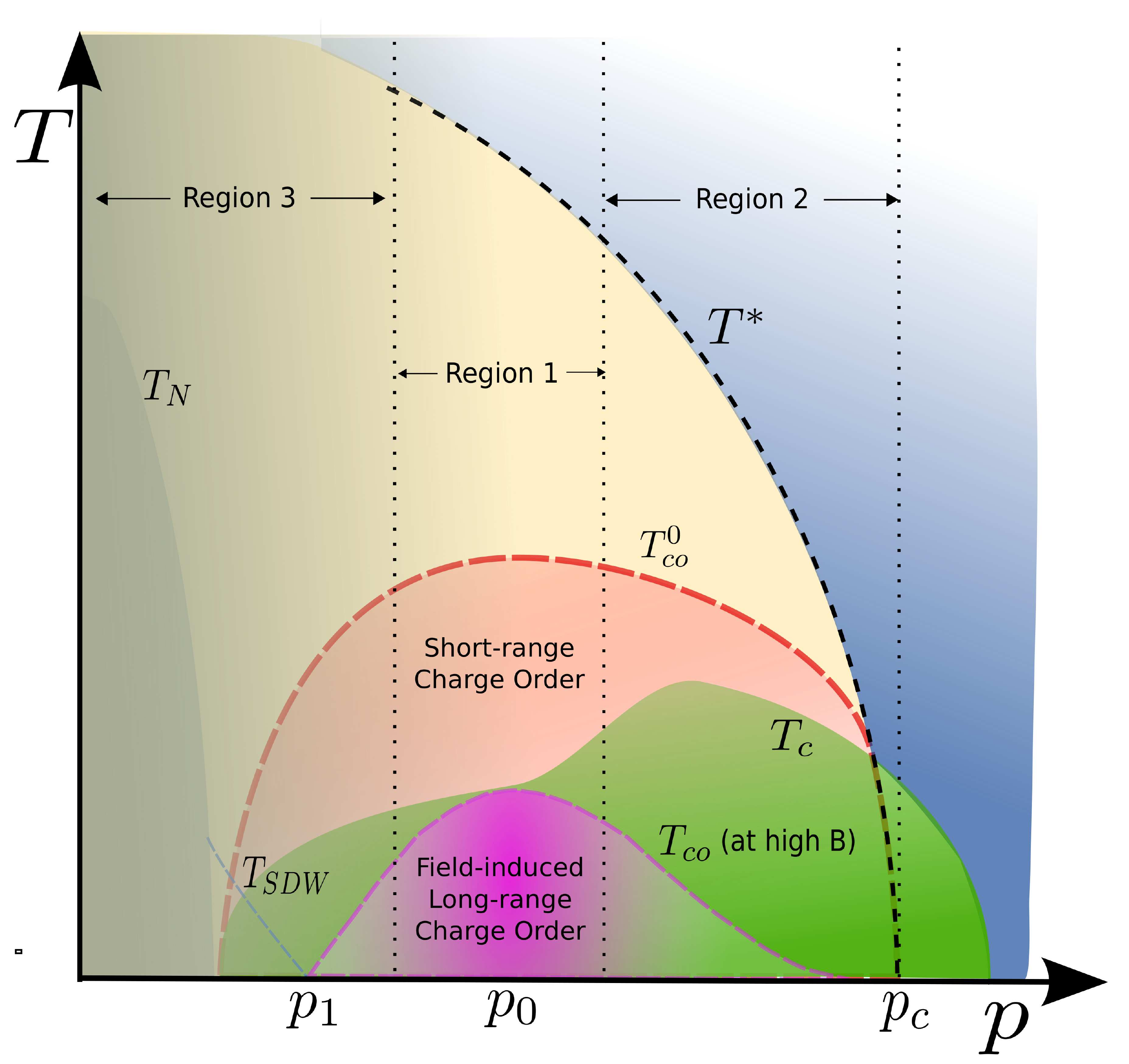}
\caption{ A schematic temperature (T)-hole doping (p) phase diagram of a typical cuprate. Cuprates show d-wave superconductivity arising from doping a Mott insulating antiferromagnet. The critical temperature $T_c$, below which the substance behave as a d-wave superconductor with an anisotropic pairing gap, forms a dome shape in the T-p phase diagram. The system shows a mysterious pseudogap phase which terminates at a temperature $T^{*}$, which is much higher than $T_c$ at low doping. This $T^*$ line approaches the doping axis with a quantum critical point ($p_c$) lying underneath the superconducting dome. The charge order in this phase diagram show two distinct behaviors: one which is short-range, observed in X-ray scattering measurements at zero or low external applied magnetic fields for $T<T_{co}^0$ and the other which is long-range, inferred from NMR or sound velocity measurements at high magnetic fields for $T<T_{co}$. In the region 1, the classical thermal fluctuations dominate resulting in $T_{co} \approx T_{c}$, as found in Eqs.~\eqref{eq:tch1} and \eqref{eq:tch2}. Even though $T_{co}^0$ is found to be greater than $T_c$, $T_{co}$ is restricted to a maximum of $T_c$, indicating a symmetry constraint between the CO and the SC. Close to $p=p_c$ (region 2), the quantum fluctuations play a key role where $T^{*} \rightarrow 0$. The region 3 signifies the low doping region where the presence of competing magnetic phases restricts the validity of the SU(2) theory described here.      
}
\label{fig:dopschem}
\end{figure}

\subsubsection*{\textbf{\textit{Schematic doping dependence}}}

Finally, we turn to the discussion of the temperature-hole doping (p) phase diagram of cuprates. In Fig.~\ref{fig:dopschem}, we sketch a schematic T-p phase diagram of a typical cuprate. We will first discuss the doping evolution of the pseudogap temperature ($T^*$). We then focus on the doping dependence of the transition temperature ($T_{co}$) of the high-field CO and contrast it with the transition temperature ($T_{co}^0$) of the short-range CO (at zero or low magnetic fields) .

First, we recall from the discussion of the NLSM that $T^*$ is proportional to the effective stiffness of the classical NLSM in Eq.~\eqref{eq:NLSM} \cite{Efetov13}. Earlier in this section, we showed that the quantum fluctuations renormalize the effective stiffness of the classical NLSM which is given as $\tilde{\rho_s}=\rho_s^0(1-g_0/g_c)$ with $g_c$ being the $T=0$ quantum critical point. We believe that $g_c$ scales with doping and is related to the $p_c$ (the pseudogap quantum critical point under the superconducting dome). The evolution of $T^*$ with doping is governed by the nature of $\tilde{\rho_s}$. As we approach $p_c$ ($\sim 0.2$), the quantum fluctuations are enhanced ($g_0 \rightarrow g_c$). The effective stiffness $\tilde{\rho}_s \rightarrow 0$ implying that the pseudogap temperature $T^* \rightarrow 0$.

We now contrast the behavior of $T_{co}$ and $T_{co}^0$. In this paper, we focus on the long-range CO transition temperature $T_{co}$ inferred from NMR or sound velocity measurements at high magnetic fields. In the pseudogap phase of Figs. \ref{fig:crg} and \ref{fig:qrg}, the system hesitates between the SC and the CO with no visible long-range order. But the pseudogap phase can accommodate a short-range CO for $T_c<T<T^*$. X-ray scattering measurements manifest the short-range CO at zero or low magnetic fields for $T<T_{co}^0$. $T_{co}^0$ is found to be greater than $T_c$ and smaller than $T^*$. Remarkably, the doping dependence of $T_{co}^0$ do not follow $T^*$. The doping dependence of $T_{co}^0$ is not the subject of this paper, but the readers may consult Ref. [\onlinecite{Morice2017skyrmions}] which explains the difference in the doping dependence of $T_{co}^0$ and $T^*$ in terms of topological defects in the SU(2) theory. Coming back to the discussion of $T_{co}$, the RG of classical NLSM gives $T_{co} \approx T_c$ as shown in Eqs.~\eqref{eq:tch1} and \eqref{eq:tch2}. We argue that this equality is valid in the region 1 (near $p= p_0 \sim 0.12$) in the T-p phase diagram (Fig.~\ref{fig:dopschem}) where the quantum fluctuations are not important. In the region 2, the effect of quantum fluctuations becomes significant. Consequently, $T^*$, $T_c$ and $T_{co}$ all approach zero. In this regime, $B_{co}$ varies quadratically with $T$ (Fig.~\ref{fig:qrg}) and $T_{co}$ becomes sensitive to the magnetic field at which the measurement is performed. If $T_{co}$ is measured at the same $B$ for all dopings, the ratio $T_{co}/T_c$ for $p \sim 0.12$ will be close to unity whereas $T_{co}/T_c$ for $p \sim 0.2$ will be much lower (as shown schematically in Fig.~\ref{fig:dopschem}). This result would encourage detailed experimental investigations of the B-T phase diagrams for $p > 0.13$.

The low doping region (region 3 in Fig.~\ref{fig:dopschem}) encounters several other competing phases like spin density wave, which are beyond the scope of this paper. But these competing magnetic phases are expected to suppress the CO making $T_{co}$ small.

\section{Conclusion}\label{sec:conc}

In this paper, we compared the competing order GL formalism and the SU(2) theory in the endeavor to explain the B-T phase diagram of the underdoped cuprates. 

We addressed the GL theory of the competing SC and CO using an effective homogeneous picture with a renormalized mass of the SC field close to the upper critical magnetic field. The strength of the repulsive interaction between the two fields decides the fate of the coexistence region in the phase diagram. A strong interaction disfavors any coexistence region. In the absence of the coexisting phase, the field corresponding to the transition from the CO phase to the SC phase is found to be temperature independent only if the temperature dependence of both the mass parameters are exactly same. We demonstrated that there is an enhanced symmetry between the SC and the CO for a range of GL parameters.

In comparison, the SU(2) theory treats the SC and the CO as two components of a composite order parameter where each of them are nearly degenerate in energy. The length of this composite order parameter remains fixed for all temperatures below the pseudogap temperature. This imposes a constraint on the SC and the CO reflecting the competition between the two. The degeneracy in the energy of the two orders is lifted by weak symmetry breaking parameters, which decide the low energy features of their competition. We showed that the presence of a coexisting phase is possible only if this symmetry is broken due to the biquadratic terms in the free energy.

In the following, we briefly highlight the features of the B-T phase diagram that distinguishes the SU(2) theory from the competing order GL formalism:

\begin{enumerate}

\item The most striking feature of the B-T phase diagram that supports the SU(2) theory is the flatness of $B_{co}$. We show that the difference in the masses of the two orders stabilizes the SC or the CO depending on the magnetic field. The transition magnetic field $B_{co}$ is exponentially suppressed (remains flat) at low temperatures due to weak thermal fluctuations. At high temperatures, the thermal fluctuations destroy any long-range order marking the pseudogap phase.  In contrast, the flat $B_{co}$ can only be achieved by extreme fine tuning of the mass parameters in the GL theory of competing orders.

\item Second demarcating feature is the similarity of $T_c$ and $T_{co}$. In an SU(2) theory, $T_c$ and $T_{co}$ are the temperatures which characterize the transition from the SC and the CO phase to the pseudogap phase respectively. Both these transitions are governed by the same energy scale (difference of the zero temperature  masses) and are thus described by similar temperatures. On the contrary, $T_c$ and $T_{co}$ in a GL theory are the temperatures where the mass of the SC and the CO fields go to zero respectively. Hence, there is no generic relation between $T_c$ and $T_{co}$ in a GL theory.    

\end{enumerate}  

In this paper, we discussed only the long-range CO and did not focus on how the long-range coherence sets in. Possible explanations for the long-range CO coherence in the SU(2) theory are very similar to the ones given in a GL theory. The constraint in the SU(2) theory will induce a finite charge order near the vortices \cite{Einenkel14} with emerging pseudospin skyrmions \cite{Morice2017skyrmions}. At low fields, these vortices are separated far from each other and the orientation of the pseudospin is fully random. If the magnetic field is increased, the vortices start to overlap and the pseudospin orientations all align in the same direction giving the long-range CO. Similar transition from the short-range CO to the long-range CO can also be explained by considering an interlayer coupling \cite{PhysRevB.92.224504}. The presence of the interlayer coupling is also expected to change the dimensionality of the CO from 2D to 3D and as a consequence, can have important consequences in its directionality.

\begin{acknowledgments}
We thank Y. Sidis for valuable discussions. This work has received financial support from the ERC, under the grant agreement AdG-694651-CHAMPAGNE.
\end{acknowledgments}

\appendix

\section{Effective free energy density of a superconductor in the presence of a magnetic field}\label{sec:appendixHC2}

In this appendix, we derive an effective homogeneous free energy density of a type-II superconductor\cite{book:17888,book:17902,RevModPhys.82.109}. Type-II superconductors possess two critical magnetic fields $B_{c1}$ (lower critical field) and $B_{c2}$ (upper critical field). For $B<B_{c1}$, the superconductor expels magnetic field completely and it is said to be in Meissner state. Once $B_{c1}<B<B_{c2}$, magnetic flux lines penetrate the sample at different locations in the sample creating vortices and the state is often termed as mixed state. If the magnetic field is further increased such that $B>B_{c2}$, the superconductivity is completely lost and the magnetic field can penetrate through the whole sample. Cuprates are well known to be extremely type-II with very small $B_{c1}$. Close to $B_{c2}$, the superconducting order parameter $\psi$ is small and can very well be described by a GL free energy. The free energy functional in the presence of magnetic field is written as\cite{book:17888,book:17902}
\begin{equation}
F_{\rm sc}-F_{\rm n}=\int \alpha'_{\psi} \psi^{2} + \frac{\beta_{\psi}}{2} \psi^4 + \frac{\lambda}{2} \left | \left ( \frac{\nabla}{i} - \frac{2e\vec{A}}{c} \right ) \psi \right |^2 dR
\label{eq:GL1}
\end{equation} 
where $\vec{A}$ is the vector potential corresponding to the magnetic field. We minimize this free energy which yields the GL equation:
\begin{equation}
\alpha'_{\psi} \psi + \beta_{\psi} \psi^3 + \frac{\lambda}{2} \left ( \frac{\nabla}{i} - \frac{2e\vec{A}}{c} \right )^2 \psi=0
\label{eq:GL2}
\end{equation}
Since the order parameter is small, we can neglect the cubic term in Eq.~\eqref{eq:GL2}. This leads to an eigen-value equation,
\begin{equation}
\frac{\lambda}{2} \left ( \frac{\nabla}{i} - \frac{2e\vec{A}}{c} \right )^2 \psi=- \alpha'_{\psi} \psi
\label{eq:GL3}
\end{equation}
If we consider the gauge $\vec{A}=(0,Bx,0)$ and an ansatz $\psi(x)=e^{ik_yy}e^{ik_zz}d(x)$, Eq.~\eqref{eq:GL3} gives,
\begin{equation}
-\frac{\partial^2 d}{\partial x^2} + \left ( \frac{2 \pi B}{\phi_0} \right )^2 \left ( x-x_0 \right )^2 d = \left ( E-k_z^2 \right ) d
\label{eq:GL4}
\end{equation} 
where $x_0=k_y \phi_0/(2 \pi B)$ and $\phi_0= h c/(2 e)$ is the flux quantum. Eq.~\eqref{eq:GL4} is similar to the Schr\"odinger equation of a charged particle in a magnetic field and the eigen-values of the equation are given by $E_n=(n+1/2)(2eB \lambda)$ for $k_z=0$ in the units $\hbar=c=1$. Near $B=B_{c2}$, $n=0$ and putting Eq.~\eqref{eq:GL4} back in Eq.~\eqref{eq:GL1}, we get
\begin{equation}
F_{\rm sc}-F_{\rm n}=\int \alpha_{\psi} \psi^{2} + \frac{\beta_{\psi}}{2} \psi^4 dR
\label{eq:GL5}
\end{equation}
where $\alpha_{\psi}=\alpha'_{\psi}+2e \lambda B$. So, for $B \approx B_{c2}$, the free energy can be written as if the order parameter is effectively homogeneous and the magnetic field just renormalizes the effective mass. The solution of Eq.~\eqref{eq:GL4} for $\psi$ is highly degenerate in $k_x$. But, away from $B_{c2}$, the non-linear term in the GL equation cannot be neglected and eigenvalue equation in Eq.~\eqref{eq:GL3} is no longer valid. Including the quartic term, it can be shown energetically that the vortices form a periodic array called Abrikosov vortex state. With the Abrikosov vortex solutions for $\psi$, we can write the free energy density as
\begin{equation}
f_{\rm sc}-f_{\rm n}=\alpha_{\psi} \bar{\psi}^{2} + \frac{\tilde{\beta_{\psi}}}{2} \bar{\psi}^4
\label{eq:GL6}
\end{equation}
where $f=F/(\int dR)$, $\int \psi^2 dR=\bar{\psi}^{2} \int dR$ and $\tilde{\beta_{\psi}}=\beta_{A} \beta_{\psi}$. $\beta_{A}$ is a parameter dependent on the geometry of the vortex lattice\cite{RevModPhys.66.1125,RevModPhys.82.109} and is found to be minimum for a triangular one with its value $\beta_{A}=1.16$.      

\section{RG equations for the classical NLSM}\label{sec:RGcNLSM}

We detail the derivation of the RG flow equations for the effective coupling constant ($t$) and the effective anisotropy parameter $\kappa$ of the NLSM in Eq.~\eqref{eq:NLSM}. Using standard techniques of RG, we integrate out the fast varying components of the free energy in Eq.~\eqref{eq:NLSM} and write an effective slow varying counterpart within one-loop order. In deriving the RG equations, we will treat $u$ as an SU(N) matrix for the sake of generality and finally analyze our results with $N=2$. We decompose $u$ as $u=u_0 \tilde{u}$, where $u_0$ and $\tilde{u}$ are the fast varying and the slow varying parts of $u$ respectively. Substituting this in Eq.~\eqref{eq:NLSM}, we get,
\begin{equation}
\frac{F}{T}=F_s^{nm}+F_s^m+F_f+F_{int}^{nm}+F_{int}^{m}
\label{eq:Frbreaking}
\end{equation}
where the superscript `nm' refers to terms arising from the gradient term in Eq.~\eqref{eq:NLSM} and `m' refers to the terms arising from the non-gradient term, the subscript `s' refers to the slow parts, `f' refers to the fast parts and `int' refers to the terms which act as interaction between the slow and the fast parts. Individual components in Eq.~\eqref{eq:Frbreaking} are given by:
\begin{equation}
F_s^{nm}=\frac{1}{t} \int tr[\nabla \tilde{u}^{\dagger} \nabla \tilde{u}]dR
\label{eq:slownm}
\end{equation}
\begin{equation}
F_s^{m}=\frac{1}{t} \int tr[\kappa \tau_3 \tilde{u}^{\dagger} \tau_3 \tilde{u}]dR
\label{eq:slowm}
\end{equation}
\begin{equation}
F_f=\frac{1}{t} \int tr[\nabla u_0^{\dagger} \nabla u_0]dR 
\label{eq:fastpart}
\end{equation}
~
\begin{equation}
F_{int}^{nm}=\frac{2}{t} \int tr[u_0^{\dagger} \nabla u_0 \tilde{u} \nabla \tilde{u}^{\dagger}]dR 
\label{eq:intnomass}
\end{equation}
\begin{equation}
F_{int}^{m}= \frac{\kappa}{t} \int tr[u_0^{\dagger}\tau_3 u_0 \tilde{u} \tau_3 \tilde{u}^{\dagger} - \tau_3]dR
\label{eq:intmass}
\end{equation}
The fast components in Eq.~\eqref{eq:Frbreaking} integrate out and do not contribute to the effective low energy free energy. On the other hand, we treat the interaction components as perturbations over the slow parts, in turn renormalizing them.  
We choose a suitable parametrization for the matrix $u_0=e^{w}$ with $w=i\sum_{a=1,N^2-1} \pi_a T^a$, where $T^a$ are the generators of SU(N) algebra and $\pi_a$ are the coefficients on the basis of $T^a$. Using this parametrization, Eq.~\eqref{eq:intnomass} becomes
\begin{equation}
F_{int}^{nm}=\frac{2}{t} \int tr[\phi_{\mu}(\nabla w w - w \nabla w)]dR
\end{equation}         
with $\phi_{\mu}=\tilde{u} \nabla \tilde{u}^{\dagger}$. Integrating over the fast variables, the contribution of the perturbation due to the $F_{int}^{m}$ up to one-loop order is $-ln(1+1/2\langle(F_{int}^{nm})^2 \rangle_w)\simeq -1/2\langle(F_{int}^{nm})^2 \rangle_w$. In momentum space, 
\begin{equation}
\langle(F_{int}^{nm})^2 \rangle_w \simeq \int_{p} \int_{q} \langle tr[A_{q,p}w_{q+p}w_{-p}]tr[A_{q',p'}w_{q'+p'}w_{-p'}] \rangle_w
\label{eq:F_1}
\end{equation} 
where $A_{p,q}=-2ip_{\mu}\phi_{\mu,-q}/t$ and we use the notations
\begin{equation}
\int_{p} \equiv \int_{\Lambda/b}^{\Lambda} \frac{d^dp}{(2\pi)^d}~{\rm and}~\int_{q} \equiv \int_{0}^{\Lambda/b} \frac{d^dq}{(2\pi)^d}
\label{eq:defint}
\end{equation}
where $\Lambda$ is the upper momentum cut-off and $1/b$ is the width of high momentum shell integrated out in every RG step and $d$ is the spatial dimension. Considering the SU(N) algebra of $w$, we obtain the following identity:
\begin{eqnarray}
&&p_{\mu}p_{\mu'} \langle tr[\phi_{\mu, q}w_{p+q} w_{-p}]tr[\phi_{\mu', q'}w_{p'+q'} w_{-p'}] \rangle_{w} \nonumber \\ 
&=& \frac{N}{4} G_{p} G_{p+q}p_{\mu}p_{\mu'}tr[\phi_{\mu,q}\phi_{\mu',-q}]
\label{eq:identity}
\end{eqnarray}  
where $G_{p}=t/p^2$ is the propagator of the fast fields $\pi^{a}$ assuming $\kappa$ to be small. Substituting this identity in Eq.~\eqref{eq:F_1}, we get
\begin{eqnarray}
\langle(F_{int}^{nm})^2 \rangle_w &\simeq& \frac{-N}{t^2} \int_{p} p_{\mu} p_{\nu} G_{p}^2 \int_{q} tr[\phi_{\mu,q}\phi_{\nu,-q}] \nonumber \\
&=& -N I \int_{q} tr[\phi_{\mu,q}\phi_{\mu,-q}]
\label{eq:F_2}
\end{eqnarray} 
where $I=\int_{p} \frac{1}{p^2}$ is the one-loop integral. Rescaling the slow fields, we can rewrite Eq.~\eqref{eq:F_2} as
\begin{equation}
\langle(F_{int}^{nm})^2 \rangle_w = -N I b^{d-2} F_s^{nm}
\label{eq:F_3}
\end{equation}  
The integral $I$ is given by:
\begin{equation}
I=\frac{\Omega^d}{(2 \pi)^{d}} \int_{\Lambda/b}^{\Lambda} dp p^{d-3}
\label{eq:oneloop_1}
\end{equation}
It diverges for $d=2$. We do an $\epsilon$ expansion around $d=2+\epsilon$ and get the integral as:
\begin{equation}
I=\frac{1}{2 \pi} lnb
\label{eq:oneloop_2}
\end{equation}
Using Eq.~\eqref{eq:F_3}, the renormalized parameter $\tilde{t}$ follows a recursion relation and can be written as 
\begin{equation}
\frac{1}{\tilde{t}}=\left \{ \frac{1}{t} - \frac{N}{4 \pi} lnb \right \} b^{\epsilon}
\label{eq:recursiont}
\end{equation} 
Taking the continuum limit and expressing $b=e^l$ where $l$ is small length rescaling in momentum, the RG differential equation for the parameter $t$ becomes:
\begin{equation}
\frac{dt}{dl}=-\epsilon t + \frac{N}{4 \pi} t^2
\label{eq:RG1}
\end{equation}
Now, we consider the perturbation term $F_{int}^{m}$ and expand with $u_0=e^{w}$. The only non-zero contribution up to order $w^2$ comes from terms: 
\begin{equation}
u_0^{\dagger}\tau_3 u_0 \tilde{u} \tau_3 \tilde{u}^{\dagger} - \tau_3= -w \tau_3 w \tilde{u} \tau_3 \tilde{u}^{\dagger}+ w^2 \tau_3 \tilde{u} \tau_3 \tilde{u}^{\dagger}
\label{eq:nmterms}
\end{equation} 
Contribution from the first term in the RHS of Eq.~\eqref{eq:nmterms} to the slow component is
\begin{eqnarray}
-\langle tr[w \tau_3 w \tilde{u} \tau_3 \tilde{u}^{\dagger}] \rangle_w &=& \sum_{a,b} \langle tr[\pi_a \pi_b T^a \tau_3 T^b \tilde{u} \tau_3 \tilde{u}^{\dagger}] \rangle_{w} \nonumber \\
&=& \sum_{a,b} \langle \pi_a \pi_b \rangle_w tr[T^a \tau_3 T^b \tilde{u} \tau_3 \tilde{u}^{\dagger}] \nonumber \\
&=& \sum_{a} \langle \pi_a \pi_a \rangle_w tr[T^a \tau_3 T^a \tilde{u} \tau_3 \tilde{u}^{\dagger}] \nonumber \\
&=& - \sum_{a} \langle \pi_a \pi_a \rangle_w \frac{1}{N} tr[\tau_3 \tilde{u} \tau_3 \tilde{u}^{\dagger}] \nonumber \\
\label{eq:mcontri_1}
\end{eqnarray} 
using the trace identity $\sum_a tr[AT^aA'T^a]=tr[A]tr[A']-(1/N) tr[AA']$, where $A$ and $A'$ are matrices of same dimension as SU(N) matrices $T^a$.
Similarly, the contribution from the second term in the RHS of Eq.~\eqref{eq:nmterms} is
\begin{equation}
\langle tr[w^2 \tau_3 \tilde{u} \tau_3 \tilde{u}^{\dagger}] \rangle_w = \left \{ N-\frac{1}{N} \right \} tr[\tau_3 \tilde{u} \tau_3 \tilde{u}^{\dagger}]
\label{eq:mcontri_2}
\end{equation}
Summing Eqs.~\eqref{eq:mcontri_1} and \eqref{eq:mcontri_2}, the whole contribution from the $F_{int}^{m}$ is given in the momentum space as 
\begin{eqnarray}
\langle F_{int}^{m} \rangle_w &=& \frac{\kappa}{t} N \int_p G_p \int_q tr[\tau_3 \tilde{u} \tau_3 \tilde{u}^{\dagger}] \nonumber \\
&=& \kappa N I \int_q tr[\tau_3 \tilde{u} \tau_3 \tilde{u}^{\dagger}] \nonumber \\
&=& \kappa \frac{N}{2 \pi} b^{2+\epsilon} F_s^{m} 
\label{eq:totalmcontri}
\end{eqnarray}
The corresponding recursion relation is given by
\begin{equation}
\frac{\tilde{\kappa}}{\tilde{t}}=\left \{ \frac{\kappa}{t} - \frac{N}{2 \pi} \kappa lnb \right \} b^{2+\epsilon}
\label{eq:recursionm}
\end{equation} 
Again taking the continuum limit, the RG differential equation for $\kappa/t$ is given by
\begin{equation}
\frac{d\left (ln\left (\frac{\kappa}{t}\right ) \right )}{dl}=-\frac{N}{2 \pi} t + 2+ \epsilon
\label{eq:RG2}
\end{equation}
In $d=2$ and for $N=2$, we have the RG equations from Eqs.~\eqref{eq:RG1} and \eqref{eq:RG2} for effective coupling constant $t$ and effective anisotropy $\kappa$ as:
\begin{equation}
\frac{dt}{dl}= \frac{1}{2 \pi} t^2
\label{eq:RG12da}
\end{equation}
\begin{equation}
\frac{d\left (ln\left (\frac{\kappa}{t}\right ) \right )}{dl}=-\frac{1}{ \pi} t + 2
\label{eq:RG22da}
\end{equation}

\section{Critical magnetic fields at $T=0$ with weak SU(2) symmetry breaking}\label{sec:BCwithcon}

In this appendix, we derive $T=0$ transition magnetic fields within a mean field picture for both quadratic and biquadratic symmetry breaking. The enhanced SU(2) symmetry between the SC and the CO is only true exactly if $\alpha'_{\phi}=\alpha'_{\psi}$ and $\gamma^2=\beta_{\psi} \beta_{\phi}$. But as explained in Sec.~\ref{sec:NLSMintro}, the constraint between the two orders is valid for all temperatures $T<T^*$. If we have a weak symmetry breaking in either the quadratic terms or in the biquadratic terms, the constraint is still valid for temperatures below $T^*$. In order to obtain the transition magnetic fields for different phases at $T=0$ but with the constraint imposed on the order parameters, we write down the GL free energy density from Eq.~\eqref{eq:GLgrad3} as: 
\begin{equation}
f[\psi,\phi]=(\alpha'_{\psi}+\zeta B) \psi^{2} + \alpha'_{\phi} \phi^2 + \frac{{\beta_{\psi}}}{2} {\psi}^4 + \frac{\beta_{\phi}}{2} \phi^4 + \gamma \psi^2 \phi^2
\label{eq:GLgrad3a}
\end{equation}
where $\alpha'_{\psi}<0$, $\alpha'_{\phi}<0$, $|\alpha'_{\psi}|>|\alpha'_{\phi}|$ and $\beta_{\psi}=\beta_{\phi}=\beta$. 

We first consider the quadratic symmetry breaking where $\gamma=\beta$, $|\alpha'_{\phi}| \ne |\alpha'_{\psi}|$ and small difference of the masses, $2\kappa_0=\alpha'_{\phi} - \alpha'_{\psi}$. Using the constraint $\psi^2+\phi^2=1$, the free energy density in Eq.~\eqref{eq:GLgrad3a} can be written in terms of the $\psi$ field only as:
\begin{equation}
f[\psi]=(-2\kappa_0+\zeta B) \psi^{2} + \frac{{\gamma}}{2}=m^{sc}_{eff} \psi^{2} + \frac{{\gamma}}{2} 
\label{eq:con1a}
\end{equation}
and in terms of the $\phi$ field only as:
\begin{equation}
f[\phi]=(2\kappa_0-\zeta B) \phi^{2} + \frac{{\gamma}}{2}=m^{co}_{eff} \phi^{2} + \frac{{\gamma}}{2} 
\label{eq:con2a}
\end{equation}   
where the effective mass of the SC field is $m^{sc}_{eff}=-2\kappa_0+\zeta B$ and the CO field is $m^{co}_{eff}=2\kappa_0-\zeta B$. The transition field $B^0$ is given by 
\begin{equation}
B^0=\frac{2\kappa_0}{\zeta}
\label{eq:conb0}
\end{equation}
If $B<B^0$, $m^{sc}_{eff}<0$ and $m^{co}_{eff}>0$ which makes only the SC phase stable. On the other hand, if $B>B^0$, $m^{sc}_{eff}>0$ and $m^{co}_{eff}<0$ which makes only the CO phase stable. There is a pseudo-spin flop first order transition at $B=B^0$, where the direction of the pseudo-spin flops from being in the SC easy plane to the CO easy plane. Thus, there cannot be any coexisting phase in the case of only quadratic symmetry breaking. 

Now, let us consider the case with both quadratic and biquadratic symmetry breaking with $\gamma \ne \beta$ and $\alpha'_{\phi} \ne \alpha'_{\psi}$. If the quadratic symmetry breaking ($2\kappa_0=\alpha'_{\phi} - \alpha'_{\psi}$) and the biquadratic symmetry breaking ($4z_{0}=\beta-\gamma$) are weak compared to the pseudogap energy scale, the constraint between the SC and the CO still holds. Applying the constraint, we can write the free energy density in Eq.~\eqref{eq:GLgrad3a} in terms of $\psi$ field only as  
\begin{eqnarray}
f[\psi]&=&(-2\kappa_0-4 z_{0}+\zeta B) \psi^{2} + 4z_{0} \psi^4 \nonumber \\  
&=&m^{sc}_{eff} \psi^{2} + 4z_{0} \psi^4
\label{eq:con3a}
\end{eqnarray}
and in terms of $\phi$ field only as:
\begin{eqnarray}
f[\phi]&=&(2\kappa_0-4 z_{0}-\zeta B) \phi^{2} + 4z_{0} \phi^4 \nonumber \\
&=&m^{co}_{eff} \phi^{2} + 4z_{0} \phi^4
\label{eq:con4a}
\end{eqnarray}
where the effective masses of the SC field and the CO field have contributions from the biquadratic symmetry breaking and are given by $m^{sc}_{eff}=-2\kappa_0-4 z_{0}+\zeta B$ and $m^{co}_{eff}=2\kappa_0-4 z_{0}-\zeta B$. If $z_0<0$ ($\gamma>\beta$), there is always a first order pseudo-spin flop transition at $B=B^0$ with no coexisting phase. If $z_0>0$ ($\gamma<\beta$), only the SC phase is stable for $B<B^0_{co}$, where $B^0_{co}$ is given by
\begin{equation}
B^0_{co}=\frac{(2\kappa_0-4z_0)}{\zeta}
\label{eq:bscconco}
\end{equation}
and only the CO phase is stable phase for $B>B^0_{sc}$, where $B^0_{sc}$
\begin{equation}
B^0_{sc}=\frac{(2\kappa_0+4z_0)}{\zeta}
\label{eq:bcoconco}
\end{equation}
In the range of fields $B^0_{co}<B<B^0_{sc}$, both the orders coexist.

\section{RG equations for quantum NLSM}\label{sec:RGqNLSM}

Renormalization group analysis of the free energy functional in Eq.~\eqref{eq:QNLSM2} is exactly similar to the one carried out in appendix \ref{sec:RGcNLSM} with a different one-loop integral $I$. In terms of the Matsubara frequencies ($\omega_n=2 \pi n/v$, with $n=0,\pm1,\pm2,..$), the one-loop integral $I$ in Eq.~\eqref{eq:oneloop_1} is  now replaced by: 
\begin{eqnarray}
I&=&\sum_{n=- \infty}^{\infty} \int_p \frac{1}{p^2+\omega_n^2} =\frac{v \Omega^d}{2 (2 \pi)^{d}} \int_{1/b}^{1} dp p^{d-2} coth\left (\frac{v}{2}p \right ) \nonumber \\
&\simeq& \frac{v \Omega^d}{2 (2 \pi)^{d}} coth\left (\frac{v}{2} \right ) \int_{1/b}^{1} dp p^{d-3} \simeq \frac{v}{4 \pi} coth\left (\frac{v}{2} \right ) lnb \nonumber \\
\label{eq:Qoneloop}
\end{eqnarray}  
The thickness of the time slab $v$ flows trivially with $l$. We write the equations in terms of a dimensionless parameter $t=T \Lambda^{d-2}/\rho_s$, which in the classical limit of $v \rightarrow 0$ maps to the coupling constant $t$ in the classical NLSM. The RG differential equations for $v$, $t$ and $\bar{\kappa}$ in $d=2$ and for $N=2$ are:  
\begin{equation}
\frac{dv}{dl}=-v
\label{eq:QRG1a}
\end{equation} 
\begin{equation}
\frac{dt}{dl}= \frac{1}{4 \pi} v t^2 coth \left ( \frac{v}{2} \right )
\label{eq:QRG2a}
\end{equation} 
\begin{equation}
\frac{d\left (ln\left (\frac{\bar{\kappa}}{t}\right ) \right )}{dl}=-\frac{t}{ 2 \pi} v coth \left ( \frac{v}{2} \right )  + 2
\label{eq:QRG3a}
\end{equation}

 \bibliographystyle{apsrev4-1}
\bibliography{Cuprates}

\end{document}